\documentclass[11pt,a4paper]{article}
\usepackage{amsmath,amssymb,amsthm}
\usepackage{hyperref}
\usepackage[margin=2.5cm]{geometry}
\usepackage{booktabs}
\usepackage{graphicx}
\usepackage{float}
\usepackage{xcolor}

\newcommand{\half}{\tfrac{1}{2}}
\newcommand{\dd}{\mathrm{d}}
\newcommand{\ee}{\mathrm{e}}
\newcommand{\Tr}{\mathrm{Tr}}

\title{Notes on Wasserstein distance and wormholes}
\author{Ville Keranen\thanks{\texttt{vkeranen1@gmail.com}}}
\date{May 2026}

\begin{document}
\maketitle

\begin{abstract}
We develop the Boltzmann--Wasserstein (BW) distance, a
temperature-dependent metric on the space of quantum theories, defined as
the optimal $W_2$ distance between Boltzmann-weighted energy spectra.
Computing it is an optimisation over wormholes: each unitary
identification of the two energy bases defines a coupling of the two
boundaries in the doubled Hilbert space, and the optimum --- the
comonotone partition function $C_{\max}$, which pairs states by rank ---
is the dominant wormhole connecting the two theories. For semiclassical
theories differing by a small entropy shift, the normalised BW distance
collapses to a squared horizon-area comparator,
$\tilde{\mathcal{W}}^2 \approx (\delta A/4G)^2/8$, with the two areas
evaluated at equal energy. When the Hamiltonians differ by an operator
$V$, the BW distance equals a long-time average of the real-time thermal two-point function of
$V$; when the thermal one-point function of $V$ vanishes --- for instance
for $V$ odd under an unbroken discrete global symmetry --- a four-point representation
appears at the next order. On the gravity side we construct the classical
saddle that computes $C_{\max}$: a Schwinger--Keldysh wormhole built from
two Euclidean caps sharing a single horizon, joined by Lorentzian
segments that adiabatically interpolate between the two theories. Its
on-shell action reproduces the spectral saddle of $C_{\max}$ --- both the
saddle-point conditions and the on-shell value --- and the Lorentzian
segments are essential: a purely Euclidean interpolation is exponentially
suppressed. The saddle captures only the rearrangement of the spectrum;
the perturbative representations retain in addition the variance of the
matrix elements of $V$, invisible to the classical geometry. We work out
two examples --- two BTZ black holes with different cosmological
constants and a $T\bar{T}$ deformation of BTZ.
\end{abstract}

\newpage
\tableofcontents
\newpage

\section{Introduction}
\label{sec:intro}

How far apart are two quantum gravity theories?  Entropic measures
like relative entropy quantify distinguishability under hypothesis
testing~\cite{Balasubramanian:2014} but are blind to the geometry of
spectral differences.  The Zamolodchikov metric provides a natural
distance along conformal manifolds~\cite{Perlmutter:2020}, but only
for exactly marginal deformations.  Circuit complexity captures the
full computational cost of mapping one theory to another, but is
notoriously hard to compute.  In this paper we develop a measure that
sits between these extremes: the Boltzmann--Wasserstein (BW)
distance,  the optimal $W_2$ distance between Boltzmann-weighted energy spectra.
For two theories with Hamiltonians $H_1, H_2$ it is defined by a
minimisation over unitaries,
\begin{equation*}
  \mathcal{W}^2(\beta)
  = \min_{U}\,\Tr\bigl(\ee^{-\beta H_1} - U\,\ee^{-\beta H_2}\,U^\dagger\bigr)^2
  = Z_1(2\beta) + Z_2(2\beta) - 2\,C_{\max}(\beta).
\end{equation*}
Expanding the square leaves the partition functions
$Z_i(2\beta) = \Tr\,\ee^{-2\beta H_i}$ together with a cross-term
$C(U) = \Tr\bigl(\ee^{-\beta H_1}\,U\,\ee^{-\beta H_2}\,U^\dagger\bigr)$,
whose maximum over $U$ is $C_{\max}$. Here the unitary $U$ identifies the
spectrum of one theory with the other, and the minimisation selects the
optimal identification. Built entirely from this partition-function data,
the $\beta$-dependence of $\mathcal{W}^2$ makes it a spectroscopic tool,
identifying the energy scale at which two theories begin to differ.

The cross-term $C_{\max}$ is where the geometry enters. In the doubled
Hilbert space, each unitary $U$ relating the two energy bases defines a
coupling of the two boundaries --- a wormhole\footnote{Throughout this paper "wormhole"
refers to the Maldacena-style two-boundary geometry of an eternal
black hole / Schwinger--Keldysh contour --- topologically a disc in
Euclidean signature, with the two asymptotic boundaries connected
by the bulk through a single horizon.  This is distinct from the
higher-topology "replica wormholes" / double-trumpet geometries
(Saad--Shenker--Stanford and follow-ups) that arise in spectral
form factors and ensemble-averaged partition functions.} --- so that computing
$\mathcal{W}^2$ is an \emph{optimisation over wormholes}: the distance is
fixed by the single pairing that maximises the combined Boltzmann weight.
The optimum, the \emph{comonotone partition function} $C_{\max}$, is the
monotone rearrangement that pairs energy eigenstates by rank. In
gravitational language this is the Brenier map $T_0(E)$, which pairs black
holes of equal horizon area, and the corresponding wormhole is the dominant
geometry connecting the two theories.

When the two theories have black holes that differ by a small entropy
shift, $\mathcal{W}^2$ collapses to a simple geometric form:
\begin{equation*}
  \tilde{\mathcal{W}}^2 \approx \frac{1}{8}\!\left(\frac{\delta A}{4G}\right)^{\!2},
\end{equation*}
where $\delta A$ is the horizon-area difference between the two black holes
\emph{at equal energy}. It is worth being explicit about which comparison
this is, as two are in play. The comonotone wormhole pairs the two theories
at equal horizon \emph{area} (the shared horizon described below), so that
the spectral difference between them is carried entirely by the Boltzmann
weights $\ee^{-\beta E}$ of the paired states. The area comparator is the
equivalent equal-\emph{energy} bookkeeping: on expanding those Boltzmann
weights, the leading difference reorganises into the entropy gap
$\delta S = \delta A/4G$ between the two theories at fixed energy, and the
three terms of $\mathcal{W}^2$ conspire into its square. The shift
$\delta S$ therefore originates in the Boltzmann weighting of the spectrum,
not in the shared horizon of the wormhole itself. In this regime the
normalised BW distance is literally a squared horizon-area comparator.

On the field-theory side, when the two Hamiltonians differ by a small
perturbation $H_2 = H_1 + \varepsilon V$, the BW distance is a
time-averaged thermal two-point function of the perturbation:
\begin{equation*}
  \mathcal{W}^2 = \varepsilon^2\beta^2\lim_{T\to\infty}\frac{1}{T}\!\int_0^T\!\!\dd t\;
  \Tr\!\bigl(V(0)\,V(t)\,\ee^{-2\beta H_1}\bigr) + O(\varepsilon^3).
\end{equation*}
In cases where the thermal one-point function of $V$ vanishes in theory 1 (e.g. due to a $\mathbb{Z}_2$ symmetry)
and the connected piece decays on the QNM
scale, the two-point answer vanishes under the time average and
the leading contribution moves to $O(\varepsilon^4)$ (unless one is working at
non-perturbative $e^{-S}$ accuracy), where it
admits a time-averaged thermal four-point representation
(Appendix~\ref{app:4pt}).

\paragraph{The Schwinger--Keldysh wormhole.}
The comonotone partition function has a classical gravitational saddle, which we construct explicitly.  The
comonotone unitary can be semiclassically implemented adiabatically along a family of
Hamiltonians interpolating between $H_1$ and $H_2$, which on the
gravity side corresponds to a Schwinger--Keldysh contour built from
two Euclidean caps sharing a single horizon and joined by a
Lorentzian segment that drives the boundary data adiabatically from
one theory to the other.  The Lorentzian part is essential: a purely Euclidean interpolation generates a
non-cancelling action that exponentially suppresses the overlap.  
The main pay-off of the construction is a precise saddle-to-saddle match.
$C_{\max}$ is defined by the saddle point of a spectral (energy) integral,
with no reference to gravity; yet the on-shell action of the SK geometry
reproduces that saddle in full --- the horizon smoothness condition
reproduces the microcanonical saddle equation
$1/\beta_1 + 1/\beta_2 = 1/\beta$, and the two Euclidean cap actions
reproduce its on-shell \emph{value}. That a genuine bulk geometry computes
the spectral saddle of $C_{\max}$ --- conditions and value alike --- is a
non-trivial check on the picture. We carry this out explicitly for two BTZ
black holes with different cosmological constants and for a $T\bar T$
deformation of BTZ, matching the gravity action against the spectral
representation of the partition functions (our meaning of ``field theory''
throughout).

\paragraph{Saddle vs.\ fluctuation.}
The semiclassical SK wormhole captures only the eigenvalue content of
$C_{\max}$: 
the rearrangement that pairs $\{E_n^{(1)}\}$ with $\{E_n^{(2)}\}$, encoded in the horizon smoothness condition and
the entropy matching. The perturbative representation retains strictly
more, and the split has a plain statistical reading: the time-averaged
two-point function computes the thermally weighted \emph{second moment}
of the diagonal matrix elements $V_{nn}$, while the saddle retains only
its squared mean $\langle V\rangle^2$. The difference --- the variance of
$V_{nn}$ about its microcanonical mean, generated by the pseudo-random
overlaps of the energy eigenvectors with $V$ --- is of order $\ee^{-S}$
and invisible to the geometry at any order in $1/S$. 
For a perturbation whose mean vanishes without a symmetry protecting the
individual matrix elements, this also sharpens the earlier statement that
the leading contribution moves to $O(\varepsilon^4)$: that holds for
$\varepsilon \gg \ee^{-S/2}$, while for smaller $\varepsilon$ the distance
is dominated by the variance term.
At the operator level the adiabatic time-ordered evolution equals
$U_{\mathrm{com}}^\dagger$ dressed by an energy-dependent dynamical
phase, together with diabatic corrections.  The phase is diagonal in
the energy basis and drops out of the BW distance identically; the
diabatic corrections make the adiabatic protocol strictly suboptimal
at any finite ramp time, and vanish only as the extent of the
adiabatic time-evolution is sent to infinity.

\paragraph{Related work.}
Optimal transport has recently appeared in the holographic context
through the emergent Wasserstein spacetime of
Hashimoto--Tanahashi~\cite{Hashimoto:2026,Hashimoto:2026b}, where the
$W_1$ distance between Husimi representations defines an emergent
radial direction.  Our approach is complementary: we work with the
$W_2$ distance between energy spectra and develop the wormhole
interpretation.  Quantum generalisations of optimal transport based
on quantum channels~\cite{DePalma:2019} provide a different
non-commutative extension.

\paragraph{Organisation.}
These notes are written in a somewhat modular fashion, so they
allow the reader to skip sections they find uninteresting.
The minimal reading is to read~\ref{sec:BW} and the
universal entropy form of~\ref{sec:he-universality}, beyond which
any single section should be more or less readable by itself.
Section~\ref{sec:BW} defines the BW distance and identifies it with
the Hoffman--Wielandt formulation of $W_2$.
Section~\ref{sec:Brenier} develops the general structure of the
Brenier map, including a Bernstein-criterion analysis of when the
overlap admits a positive single-wormhole decomposition.
Section~\ref{sec:semiclassical} constructs the Schwinger--Keldysh
wormhole and works out the BTZ and $T\bar T$ examples.
Section~\ref{sec:area-comparator} derives the area comparator, and
Section~\ref{sec:perturbative} develops the two-point function
representation.  Appendix~\ref{app:slow-roll} gives the slow-roll
construction of the Lorentzian segment; Appendix~\ref{app:fzzt}
works out a single eigenstate perturbation via FZZT branes in
JT gravity; 
Appendix~\ref{app:4pt}
records the four-point representation of $\mathcal W^2$ at
$O(\varepsilon^4)$ for perturbations with vanishing thermal one-point
function.

\paragraph*{Note on this work.}
This paper was written part-time, outside my main professional
activity (quantitative finance and machine learning, where $W_2$ is
a standard tool); the wormhole picture is one I had wanted to write
up for some time.  The exposition, numerical calculations, and most
of the figures were produced with substantial AI assistance.
Responsibility for any errors is mine, and corrections are welcome.
All the results (except Appendix A) have been calculated with the
good old pen and paper method. Appendix A on the other hand
has been worked out with AI written mathematica notebooks.
My literature search is necessarily incomplete given my distance from
the field, and I apologise in advance for relevant prior work not
cited.

\section{The Boltzmann--Wasserstein distance}
\label{sec:BW}

\subsection{Optimal transport and the Wasserstein distance}

The Wasserstein distance between two probability distributions $\mu$
and $\nu$ on $\mathbb{R}$ is defined by finding the cheapest way to
rearrange one into the other:
\begin{equation}\label{eq:W2-OT}
  W_2^2(\mu,\nu) = \inf_{\gamma \in \Gamma(\mu,\nu)}
  \int |x - y|^2\,\dd\gamma(x,y),
\end{equation}
where $\Gamma(\mu,\nu)$ is the set of all couplings --- joint
distributions whose marginals are $\mu$ and $\nu$.  For
one-dimensional distributions the optimum is the monotone
rearrangement: pair the $k$-th smallest element of $\mu$ with the
$k$-th smallest of $\nu$~\cite{Villani:2009,Brenier:1991}.

In this work we consider the Wasserstein distance between the spectral
densities (density of energy eigenstates) of two theories. One immediate
issue is that the density of states of any system relevant to gravity
increases indefinitely with energy, leading to a non-normalisable density.
We introduce a natural regularisation for this, by considering a cost function
constructed from Boltzmann weights instead of using a simple squared distance
between energies.

\subsection{Definition}

Consider two quantum systems with Hamiltonians $H_1$ and $H_2$.
The \emph{Boltzmann--Wasserstein distance} at inverse temperature
$\beta$ is
\begin{equation}\label{eq:BW-def}
  \mathcal{W}^2(\beta) \equiv \min_{U \in U(N)}
  \Tr\Bigl(\ee^{-\beta H_1} - U\,\ee^{-\beta H_2}\,U^\dagger
  \Bigr)^2.
\end{equation}
Unlike the standard Wasserstein distance applied to normalised
spectral densities, we work with unnormalised Boltzmann-weighted
measures.  The Boltzmann weight $\ee^{-\beta E}$ serves as a natural
regulator for the non-normalisable gravitational density of states,
and the temperature $\beta$ controls which part of the spectrum is
being compared.
That the unitary minimisation in~\eqref{eq:BW-def} does yield a
Wasserstein distance is the content of the Hoffman--Wielandt
inequality~\cite{HoffmanWielandt:1953}, which identifies
$\inf_U \|A - UBU^\dagger\|_{\mathrm{HS}}^2$ with the $W_2^2$
distance between the eigenvalue
measures of $A$ and $B$; see~\cite{Jacelon:2018,Jacelon:2026} for
generalisations to C$^*$-algebras and II$_1$ factors.

\begin{figure}[t]
\centering
\includegraphics[width=\textwidth]{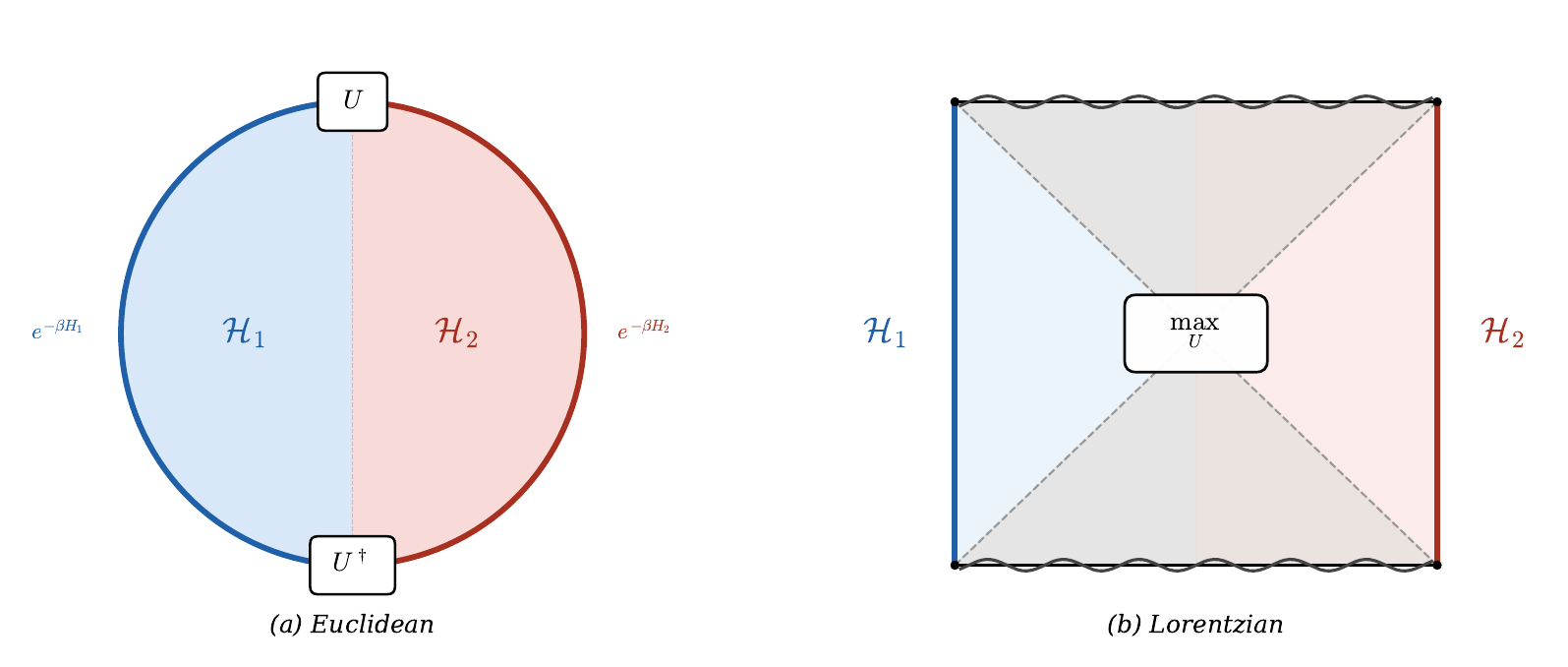}
\caption{Wormhole interpretation of the cross-term $C(U)$.
(a)~Euclidean picture: the disc is cut into two halves.  The left
boundary (blue) corresponds to $\ee^{-\beta H_1}$ of theory~1,
the right boundary (red) to $\ee^{-\beta H_2}$ of
theory~2.  The unitary~$U$ glues the two half-discs at the top and
bottom junctions.
(b)~Lorentzian continuation: a two-sided wormhole, with the left
asymptotic boundary governed by $\mathcal{H}_1$ and the right
by $\mathcal{H}_2$.  The BW distance selects the $U$ that maximises
the overlap.}
\label{fig:wormhole}
\end{figure}

Expanding the square, the minimisation reduces to maximising the
cross-term
\begin{equation}\label{eq:C-def}
  C(U) = \Tr\!\bigl(\ee^{-\beta H_1}\,U\,\ee^{-\beta H_2}\,
  U^\dagger\bigr) = \sum_{i,j} \ee^{-\beta( E_i^{(1)}
  + E_j^{(2)})} |U_{ij}|^2,
\end{equation}
which has a direct wormhole interpretation.  To see this, work in the
doubled Hilbert space $\mathcal{H}_1 \otimes \mathcal{H}_2$ and use
cyclicity to write
\begin{align}\label{eq:C-TFD}
C(U) &= \Tr\!\bigl(\ee^{-\beta H_1/2}\,U\,\ee^{-\beta H_2/2}\,
\ee^{-\beta H_2/2}\,U^\dagger\,\ee^{-\beta H_1/2}\bigr)\nonumber
\\
&= \sum_{i,j}\sum_{i',j'}
\ee^{-\beta E_i^{(1)}/2 - \beta E_j^{(2)}/2}\,U_{ij}\;
\ee^{-\beta E_{i'}^{(1)}/2 - \beta E_{j'}^{(2)}/2}\,U_{i'j'}^*\;
\delta_{ii'}\delta_{jj'}\nonumber
\\
&= \bigl|\!\sum_{ij}\ee^{-\beta E_i^{(1)}/2-\beta E_j^{(2)}/2}
\,U_{ij}\,|E_i^{(1)}\rangle |E_j^{(2)}\rangle\bigr|^2
= \langle U|U\rangle\,,
\end{align}
where in the second line we evaluated the trace by inserting complete
sets of energy eigenstates $\{|E_i^{(1)}\rangle\}$ and
$\{|E_j^{(2)}\rangle\}$, and in the third line we recognised the
result as the squared norm of a state in
$\mathcal{H}_1\otimes\mathcal{H}_2$.  The state
\begin{equation}\label{eq:U-TFD}
  |U\rangle = \sum_{ij}\ee^{-\beta E_i^{(1)}/2-\beta E_j^{(2)}/2}
  \,U_{ij}\,|E_i^{(1)}\rangle |E_j^{(2)}\rangle
\end{equation}
is a generalisation of the thermofield double: for $H_1 = H_2$ and
$U = \mathbf{1}$ it reduces to the standard TFD state dual to the
eternal wormhole~\cite{Maldacena:2001}, while a non-trivial $U$
implements a different pairing of the two sides
(Figure~\ref{fig:wormhole}).

\subsection{The comonotone solution and the Brenier map}

The maximum of $C(U)$ is achieved by the \emph{comonotone} unitary
that pairs eigenstates by rank:
\begin{equation}\label{eq:Cmax}
  C_{\max} = \sum_{n=1}^{\min(N_1,N_2)} \ee^{-\beta( E_n^{(1)}
  + E_n^{(2)})},
\end{equation}
where $E_{n+1}^{(i)} \ge E_{n}^{(i)}$ are ordered eigenvalues. This result can be understood
in two steps.

\paragraph{Local step: Lagrange multipliers.}
Write $A = \mathrm{diag}(\ee^{-\beta E_i^{(1)}})$ and
$B = \mathrm{diag}(\ee^{-\beta E_j^{(2)}})$, so that
$C(U) = \Tr(AUBU^\dagger)$.  Impose the unitarity constraint
$U^\dagger U = \mathbf{1}$ with a Hermitian Lagrange multiplier
$\Lambda$:
\begin{equation*}
  \mathcal{L}(U,U^\dagger,\Lambda)
  = \Tr(AUBU^\dagger)
  - \Tr\bigl(\Lambda(U^\dagger U - \mathbf{1})\bigr).
\end{equation*}
Varying with respect to $U^\dagger$ (treating $U$ and $U^\dagger$ as
independent) gives the stationarity condition $AUB = U\Lambda$, which
after multiplication by $U^\dagger$ on the left reads
$U^\dagger A U B = \Lambda$.  Hermiticity of $\Lambda$ then forces
\begin{equation}\label{eq:comonotone-commutator}
  [\,U^\dagger A U,\,B\,] = 0\,,
\end{equation}
so $U^\dagger A U$ and $B$ must commute.  When $B$ has non-degenerate
spectrum this means $U^\dagger A U$ is diagonal in the standard
basis, and since $A$ is also diagonal, $U$ must be a permutation
matrix up to an arbitrary diagonal phase matrix (the phases drop out
of $|U_{ij}|^2$ and so do not affect $C(U)$).  Labelling permutations
so that $U_\sigma|i\rangle = |\sigma(i)\rangle$, the critical values
are $C_\sigma = \sum_i a_i b_{\sigma(i)}$ where
$a_i = \ee^{-\beta E_i^{(1)}}$ and $b_j = \ee^{-\beta E_j^{(2)}}$.
The Lagrange multiplier at any such critical point is the diagonal
pair-weight matrix,
$\Lambda_{ii} = a_i b_{\sigma(i)} = \ee^{-\beta(E_i^{(1)} +
E_{\sigma(i)}^{(2)})}$.  Degeneracies in $A$ or $B$ relax the
diagonal condition: within a degenerate subspace any unitary rotation
leaves $C$ invariant, which is why $C_{\max}$ is unique but the
maximising unitary is not.

\paragraph{Global step: pair-swap inequality.}
Among the $n!$ permutations, which maximises $C_\sigma$?  
As an example, consider the first two eigenvalues of theory 1 $E_1^{(1)}<E_2^{(1)}$ and
theory 2 $E_1^{(2)}<E_2^{(2)}$, which we for concreteness assume to be non-degenerate. 
There are two pairings differing by one permutation that contribute
to $C$
\begin{align}
C_{\rm com} &= e^{-\beta (E_1^{(1)} + E_1^{(2)})} + e^{-\beta (E_2^{(1)} + E_2^{(2)})} + \ldots\quad &(\textrm{ordered})
\\
C_{\rm swap} &= e^{-\beta (E_1^{(1)} + E_2^{(2)})} + e^{-\beta (E_2^{(1)} + E_1^{(2)})} + \ldots\quad &(\textrm{swapped}).
\end{align}
It is easy to see that $C_{\rm com} > C_{\rm swap}$ since
\begin{equation}
C_{\rm com} - C_{\rm swap} = \Big(e^{-\beta E_1^{(1)}} - e^{-\beta E_2^{(1)}}\Big)\Big(e^{-\beta E_1^{(2)}} - e^{-\beta E_2^{(2)}}\Big) > 0.
\end{equation}
Thus, swapping any pair of eigenvalues out of rank order decreases
$C$, so the global maximum is the rank-matched (comonotone)
permutation: pair the lightest with the lightest, the next-lightest
with the next-lightest, and so on.  Note that the rearrangement is on
the Boltzmann weights, whose order is reversed relative to the
energies (lowest $E \leftrightarrow$ largest $\ee^{-\beta E}$); the
comonotone pairing thus matches largest weight with largest weight.
This is the temperature-dressed form of the von~Neumann trace
inequality $\Tr(AB) \le \sum_i a_i^{\downarrow} b_i^{\downarrow}$ over
unitarily-related pairs.

When $N_1 \neq N_2$, we extend the smaller spectrum by formal
infinite-energy states (zero Boltzmann weight), so that the
sum~\eqref{eq:Cmax} runs over $\min(N_1,N_2)$ paired terms and the
unpaired states of the larger theory contribute to $Z_i(2\beta)$
only.

In the continuum,\footnote{In the case of matrix models (like JT) at finite $N$ the continuum
BW distance is only approximate, and the original eigenvalue form should be used instead.} the sum over states~\eqref{eq:Cmax} becomes an
integral over a continuous rank index $n$.  For each theory define
the (unnormalised) cumulative distribution function
\begin{equation}
  n_i(E) = \int_0^E \rho_i(E')\,\dd E',
\end{equation}
where $\rho_i(E)$ is the density of states of theory~$i$.  The
comonotone pairing identifies eigenvalues with equal rank,
$n_1(E_1) = n_2(E_2) \equiv n$, defining inverse functions
$E_i(n) = n_i^{-1}(n)$.  The discrete sum becomes
\begin{equation}
  C_{\max} = \sum_{n=1}^{\min(N_1,N_2)} \ee^{-\beta(E_n^{(1)}
  + E_n^{(2)})} \;\rightarrow\; \int_0^{\infty}\!\dd n\;
  e^{- \beta E_1(n) - \beta E_2(n)}.
\end{equation}
Changing variables from $n$ to $E_1$ via $\dd n = \rho_1(E_1)\,\dd E_1$ gives
\begin{equation}\label{eq:ZBr-energy}
  C_{\max} = \int_0^{\infty}dE_1\, \rho_1(E_1)\, e^{- \beta E_1 - \beta T_0(E_1)} = \Tr \,e^{- \beta H_1 - \beta T_0(H_1)} ,
 \end{equation}
 where we have introduced the Brenier map $T_0(E)$ as the energy in theory 2
 below which it has the same number of states as the theory 1 has below $E$ i.e.
\begin{equation}\label{eq:Brenier-practical}
n_1(E) = \int_0^E \rho_1(E')\,\dd E'
  = \int_0^{T_0(E)} \rho_2(E')\,\dd E' = n_2(T_0(E)).
\end{equation}
In the form~\eqref{eq:ZBr-energy} the right-hand side is manifestly a
single thermal trace, with $H_1$ shifted by $T_0(H_1)$ --- the
deformation of theory 2's spectrum into theory 1's basis under the
rank-pairing.  We will refer to $C_{\max}$ as the \emph{comonotone
partition function} throughout: ``comonotone'' to signal the
rank-pairing of eigenvalues that defines it, and ``partition
function'' to signal that it sums Boltzmann factors as in
$Z(\beta)$.  The BW distance is then
\begin{equation}\label{eq:BW-result}
  \mathcal{W}^2(\beta) = Z_1(2\beta) + Z_2(2\beta) - 2\,C_{\max}(\beta).
\end{equation}
While for most of our calculations, we study the unnormalised quantity, it is
also convenient to define the normalised Boltzmann-Wasserstein distance that is
between 0 and 1
\begin{equation}
\tilde{\mathcal{W}}^2(\beta)  =  \frac{\mathcal{W}^2(\beta)}{Z_1(2\beta) + Z_2(2\beta) }.
\end{equation}

$\mathcal{W}(\beta)$ is a genuine metric on the space of spectra.
Expanding~\eqref{eq:BW-result} using the comonotone~\eqref{eq:Cmax}
gives
\begin{equation}\label{eq:W-ell2}
  \mathcal{W}^2(\beta) = \sum_{n=1}^{N}
  \bigl(\ee^{-\beta E_n^{(1)}} - \ee^{-\beta E_n^{(2)}}\bigr)^2,
\end{equation}
so $\mathcal{W}(\beta) = \|\mathbf{a} - \mathbf{b}\|_{\ell^2}$ with
$a_k = \ee^{-\beta E_k^{(1)}}$ and $b_k = \ee^{-\beta E_k^{(2)}}$ the
Boltzmann weights of the two theories (sorted in decreasing order).
The BW distance inherits the metric property from the vector $\ell^2$ norm.

As an illustration of the role of $U$ we can contrast two
limiting cases. 
The comonotone $U$ corresponds to the minimum free-energy wormhole --- the
dominant connected geometry linking the two theories. At the other end of
the spectrum is a Haar-random $U$,
for which a typical realisation gives
$C(U_{\rm Haar}) \approx Z_1(\beta)Z_2(\beta)/N$ (with $N$ the
Hilbert-space dimension), so the boundaries decouple --- no wormhole
connects them.  Different choices of $U$ thus determine the
wormhole's properties, including whether it exists at all.

\section{General structure of the Brenier map}
\label{sec:Brenier}

\subsection{High-energy universality}
\label{sec:he-universality}

At high energies, where $\rho_i(E) = \ee^{S_i(E)}$ grows rapidly,
the CDF is dominated by its upper endpoint
\begin{equation}
n_i(E) = \int_0^E dE' e^{S_i(E')} \sim e^{S_i(E)}.
\end{equation}
The Brenier condition
$n_1(E) = n_2(T_0(E))$ then reduces to entropy matching:
\begin{equation}\label{eq:entropy-matching}
  S_1(E) = S_2(T_0(E)) + O(1),
\end{equation}
and differentiating:
\begin{equation}\label{eq:Tprime-universal}
  T_0'(E) = \frac{S_1'(E)}{S_2'(T_0(E))}.
\end{equation}
The UV Brenier map is determined entirely by the entropy functions.
In gravitational language, $S = A/(4G)$, so the Brenier map pairs
black holes of equal horizon area.

\subsection{Complex partition function decomposition}
\label{sec:wormhole-decomp}

The comonotone partition function~\eqref{eq:ZBr-energy} involves the nonlinear
Boltzmann factor $\ee^{-\beta T_0(E)}$, which prevents direct
evaluation as a partition function of theory~1 alone.  We can unfold
it by inserting the identity
\begin{equation}\label{eq:Bromwich-identity}
  \ee^{-\beta T_0(E)} = \frac{1}{2\pi i}
  \int_{c-i\infty}^{c+i\infty} \dd\sigma\;
  \ee^{\sigma E}\;\widetilde{K}(\sigma;\beta),
\end{equation}
where $\widetilde{K}(\sigma;\beta)$ is the Laplace transform
of $\ee^{-\beta T_0(E)}$ with respect to $E$.  Substituting
into~\eqref{eq:ZBr-energy} and exchanging the $E$ and $\sigma$
integrals, the $E$-integral becomes $\int \dd E\,\rho_1(E)\,
\ee^{-(\beta - \sigma)E} = Z_1(\beta - \sigma)$, giving
\begin{equation}\label{eq:ZBr-subordination}
  C_{\max}(\beta) = \frac{1}{2\pi i}
  \int_{c-i\infty}^{c+i\infty} \dd\sigma\;
  \widetilde{K}(\sigma;\beta)\,Z_1(\beta - \sigma).
\end{equation}
Each factor $Z_1(\beta - \sigma)$ is a partition function of theory~1
at complex temperature $\beta - \sigma$ --- gravitationally, a
wormhole geometry with complex modular parameter.  The kernel
$\widetilde{K}$ encodes all information about theory~2 through the
Brenier map $T_0$.

\paragraph{Wormhole interpretation.}
The contour in~\eqref{eq:ZBr-subordination} runs parallel to the
imaginary axis: $\sigma = \sigma_R + i\sigma_I$ with $\sigma_R$ fixed
and $\sigma_I \in (-\infty,\infty)$.  Writing $\beta' = \beta - \sigma_R$
and $t = -\sigma_I$, the factor
$Z_1(\beta' + it)$ has a clean gravity interpretation as a
time-evolved thermofield double:
\begin{align}
  Z_1(\beta'-it)
  &= \Tr\,\ee^{-\beta' H/2}\,\ee^{-iHt}\,\ee^{-\beta' H/2}
  = \langle \beta'|\,\ee^{-iHt}\,|\beta'\rangle\,,
\end{align}
with $|\beta'\rangle = \sum_j \ee^{-\beta' E_j/2}|E_j\rangle|E_j\rangle$
the Hartle--Hawking state of the wormhole prepared by a Euclidean cap
of length~$\beta'$.  These are the time-evolved
wormholes studied e.g. by Hartman and
Maldacena~\cite{Hartman:2013}: both asymptotic boundaries are evolved
by $\ee^{-iHt/2}$ in the \emph{same} direction, so the Lorentzian
time~$t$ controls the growth of the wormhole interior, with $t = 0$
being the static Euclidean geometry and $|t| > 0$ producing
progressively longer wormholes.  The
decomposition~\eqref{eq:ZBr-subordination} therefore represents
$C_{\max}$ as a continuous family of complex-temperature
wormholes weighted by the kernel~$\widetilde K(\sigma;\beta)$.  Note
that this contour is a complex deformation of the Euclidean thermal
cigar, not a Schwinger--Keldysh contour.

This representation is not quite a single-theory quantity: the
dependence on theory~2 is hidden in the kernel $\widetilde{K}$, which
knows about the Brenier map and hence about $\rho_2$.

\subsection{The Bernstein criterion: when is the decomposition positive?}
\label{sec:bernstein}

A natural question is whether the kernel $\widetilde{K}$ can be
chosen as a positive measure, so that $C_{\max}$ is a positive
superposition of partition functions at real temperatures.  If so, the
complex partition function decomposition would have a direct statistical interpretation:
$C_{\max}$ as a thermal average over an ensemble of geometries.

The answer is controlled by classical results of
Bernstein~\cite{Bernstein:1929,Widder:1941,Schilling:2012}.  A positive representation
\begin{equation}\label{eq:positive-mixture}
  \ee^{-\beta T_0(E)} = \int_0^\infty \mu(\dd s)\;\ee^{-s E},
  \qquad \mu \geq 0
\end{equation}
exists if and only if $\ee^{-\beta T_0(E)}$ is \emph{completely
monotone} in $E$ --- meaning its derivatives alternate in sign:
$f \geq 0$, $f' \leq 0$, $f'' \geq 0$, and so on.  Requiring this
to hold uniformly in $\beta > 0$ is equivalent, by the composition
theorem for Bernstein functions, to $T_0$ being a \emph{Bernstein
function}: $T_0 \geq 0$ and $T_0'$ is completely monotone, i.e.,
\begin{equation}\label{eq:bernstein-def}
  (-1)^{n-1} T_0^{(n)}(E) \geq 0 \quad \text{for all } n \geq 1.
\end{equation}
Concretely: $T_0' \geq 0$ (monotone), $T_0'' \leq 0$ (concave),
$T_0''' \geq 0$, and so on.  At any \emph{fixed} $\beta$ the Bernstein
condition on $T_0$ is sufficient but not strictly necessary for
positivity of $\mu$ at that $\beta$; however, the failure modes
relevant to our examples (Brenier maps with both concave and convex
regions, generated by localised spectral perturbations) violate the
complete-monotonicity of $\ee^{-\beta T_0}$ at small $\beta$ and so
fail uniformly.  The physical intuition is that a Brenier
map whose derivative $T_0'(E) = \rho_1(E)/\rho_2(T_0(E))$ decreases
smoothly and monotonically can be unfolded into a positive mixture of
linear maps $E \mapsto sE$, each corresponding to a definite inverse
temperature.  When $T_0'$ is not monotonically decreasing --- for
instance, when it dips and then recovers --- this unfolding requires
negative or complex weights, i.e.\ partition functions at complex
temperatures.

\paragraph{Linear Brenier maps.}
The simplest example is the two-BTZ case of
Section~\ref{sec:btz}, where $T_0(E) = \alpha E$ with
$\alpha = \ell_1/\ell_2$.  This is
trivially Bernstein (all higher derivatives vanish), and the kernel is
$\mu = \delta(s - \alpha\beta)$: a single partition function at
effective temperature $\beta(1 + \alpha) = \beta_{\rm eff}$,
recovering~\eqref{eq:como-btz}.

\begin{figure}[t]
\centering
\includegraphics[width=\textwidth]{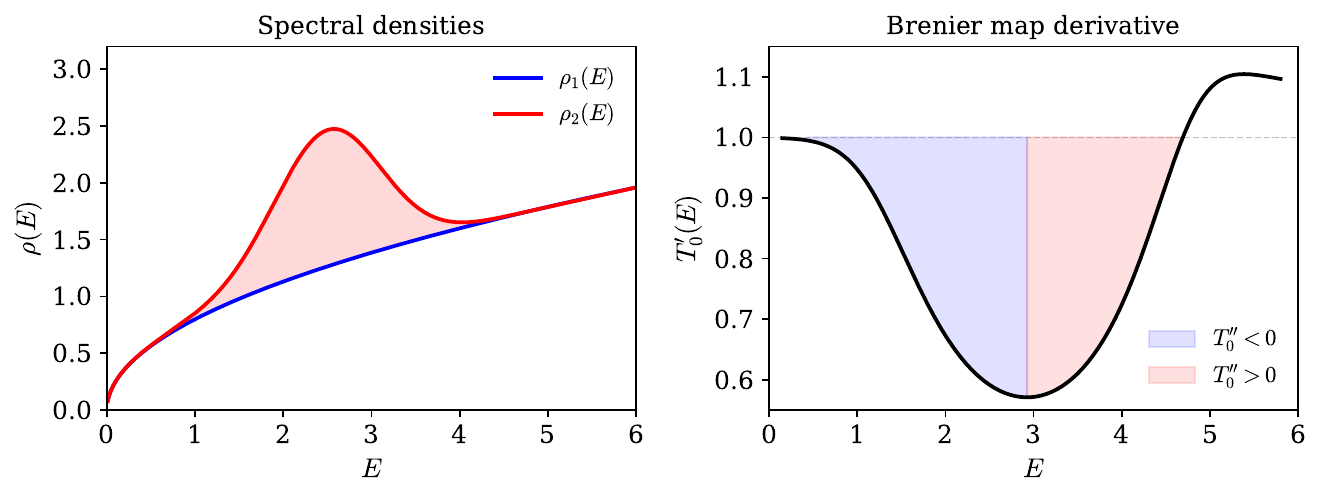}
\caption{Effect of a localised perturbation on the Bernstein
criterion.  Left: spectral densities $\rho_1(E)$ (blue) and
$\rho_2 = \rho_1 + \delta\rho$ (red), where $\delta\rho$ is
concentrated in a narrow energy window.  Right: the derivative
$T_0'(E) = \rho_1(E)/\rho_2(T_0(E))$ dips below~$1$ near the
perturbation and recovers, producing both concave ($T_0'' < 0$)
and convex ($T_0'' > 0$) regions.  The convex region violates
the Bernstein condition, so the complex partition function decomposition necessarily
requires complex temperatures.}
\label{fig:bernstein}
\end{figure}

\paragraph{Generic failure.}
In practice, the Bernstein condition is restrictive.  The
weakest necessary condition is concavity: $T_0''(E) \leq 0$ for all
$E > 0$.  Since $T_0'(E) = \rho_1(E)/\rho_2(T_0(E))$, concavity
requires the density ratio to be a monotonically decreasing function
of energy.  Any perturbation that adds states in a localised energy
window produces a Brenier map with both concave and convex regions,
immediately violating the criterion
(Figure~\ref{fig:bernstein}).

\paragraph{Subleading entropy obstruction}
The failure is generic for theories with the same leading
entropy.  If $\rho_2(E) \geq \rho_1(E)$ for all $E$ and both
theories share the same leading growth
($\rho_1(E)/\rho_2(E) \to 1$ as $E \to \infty$), then $T_0$ cannot
be a Bernstein function.  The argument is simple: since
$\rho_2 \geq \rho_1$, the Brenier map satisfies $T_0(E) \leq E$, so
$T_0'$ is on average less than~$1$.  But
$T_0'(E) \to \rho_1/\rho_2 \to 1$ at high energies.  A function
that dips below~$1$ and returns to~$1$ must increase somewhere,
giving $T_0'' > 0$ and violating concavity. 

When Bernstein fails --- generically, for any spectral perturbation
that adds states in a localised window --- the
kernel~$\widetilde K$ is not a positive measure and no single
black hole/wormhole from~\eqref{eq:ZBr-subordination} dominates: the
integral instead builds up gradually over a continuum of complex
temperatures. 

\section{The Schwinger--Keldysh wormhole}
\label{sec:semiclassical}

When the two theories differ at the thermodynamic level ---
different central charges or a
$T\bar{T}$ deformation --- the fractional spectral difference
$\delta\rho/\rho$ is $O(1)$ and the comonotone partition function has a
semiclassical wormhole saddle.

\subsection{The wormhole saddle point}
\label{sec:saddle}

The comonotone partition function
$C_{\max} = \int \dd E\,\rho_1(E)\,\ee^{-\beta(E + T_0(E))}$
is dominated by a saddle at $E = E_*$ where
\begin{equation}\label{eq:saddle-condition}
  S_1'(E_*) = \beta(1 + T_0'(E_*)).
\end{equation}
At the saddle, the Brenier condition $n_1(E_*) = n_2(T_0(E_*))$
reduces at leading exponential order to $S_1(E_*) = S_2(T_0(E_*))
\equiv S$ (cf.~\eqref{eq:entropy-matching}): both theories share the
same entropy up to $O(1)$ corrections. The on-shell value is
\begin{equation}\label{eq:C-gravity}
 \ln C_{\max} \approx -\beta(E_1 + E_2) + S,
\end{equation}
where $E_1 = E_*$, $E_2 = T_0(E_*)$, and $S$ is the common entropy.
The structure is suggestive: two boundary contributions and a shared
entropy, reminiscent of a wormhole with two caps sharing a horizon
of area $A = 4GS$.

Taking a derivative of the entropy 
matching condition gives $T_0'(E) = S_1'(E)/S_2'(T_0(E))$ which gives the saddle 
condition
\begin{equation}\label{eq:saddle-condition-2}
  S_1'(E_*) = \beta\Big(1 + \frac{S_1'(E_*)}{S_2'(T_0(E_*))}\Big).
\end{equation}
Using the microcanonical inverse temperatures
$\beta_i = S_i'(E_i)$, the saddle-point
condition~\eqref{eq:saddle-condition-2} takes the form
\begin{equation}\label{eq:Tmc-sum}
 \frac{1}{\beta_1} + \frac{1}{\beta_2} 
  = \frac{1}{\beta}\,.
\end{equation}
Defining the Euclidean angle fractions
$\Theta_i = 2\pi\beta/\beta_i$, this becomes
$\Theta_1 + \Theta_2 = 2\pi$, which is reminiscent of a smoothness
condition for a Euclidean geometry with two caps.

At this stage the gravitational interpretation is not fixed.  The
saddle of the $E$-integral exists whenever the integrand has a smooth
maximum, but we have not been able to identify it with the saddle of
a single gravitational theory, and we leave the existence of such a
single-boundary construction as an open question.  In what follows we
adopt instead a manifestly two-boundary realisation --- the
Schwinger--Keldysh contour described next --- which makes both
theories explicit on equal footing and naturally accommodates the
saddle-point condition above.

\subsection{Example: two BTZ black holes}
\label{sec:btz}

We begin with the simplest example, which already exhibits all the
key features and provides the gravitational realisation of the
spectral saddle.

\paragraph{Conventions.}
To compare two theories with different cosmological constants we put
them on a common boundary cylinder.  We do this by writing the BTZ
metric~\cite{BTZ:1992} of theory~$i$ as
\begin{equation}\label{eq:btz-common}
  \dd s^2 = -(r^2 - r_+^2)\,\dd t^2
  + \frac{\ell_i^2}{r^2 - r_+^2}\,\dd r^2
  + r^2\,\dd\phi^2,
  \qquad \phi \sim \phi + 2\pi.
\end{equation}
Stripping the leading $r^2$ at infinity gives boundary metric
$g^{(0)} = -\dd t^2 + \dd\phi^2$ for both theories, with proper
circumference $L = 2\pi$ independent of~$\ell_i$.  (This is BTZ in the
boundary-time coordinate where the asymptotic metric is
$\ell$-independent; the more familiar form
$g_{tt} = -(r^2-r_+^2)/\ell^2$, $g_{\phi\phi} = r^2$ is recovered by
$t \to t/\ell$.)  In this
common frame Brown--Henneaux gives $c_i = 3\ell_i/(2G)$, and on BTZ
\begin{equation}\label{eq:E-common-frame}
  E_i = \frac{r_+^2}{8G\,\ell_i},\quad
  \beta_i^H = \frac{2\pi\ell_i}{r_+},\quad
  S = \frac{\pi r_+}{2G},\quad
  E_{{\rm vac},i} = -\frac{\ell_i}{8G},
\end{equation}
with $E_i = \int_0^{2\pi}\dd\phi\,T^{t}{}_t^{(\text{hol})}$ the
Balasubramanian--Kraus energy and $I_i = \beta_i^H E_i - S$ the
on-shell Euclidean action at the Hawking saddle.

Take two 2+1 gravity theories with AdS radii $\ell_1$ and $\ell_2$.
Both have Cardy density
$\rho_i(E) \sim \ee^{2\pi\sqrt{\ell_i E/(2G)}}$,
and the Brenier map follows from entropy matching~\eqref{eq:entropy-matching}:
\begin{equation}\label{eq:brenier-btz}
  T_0(E) = \frac{\ell_1}{\ell_2}\, E.
\end{equation}
The map is linear because the energy--entropy relation
$E = 3S^2/(4\pi^2 c)$ is separable: $E(\ell, S) = f(\ell)\,g(S)$.
The comonotone partition function collapses to a single partition function at
effective temperature:
\begin{equation}\label{eq:como-btz}
  C_{\max} = Z_1(\beta_{\rm eff}), \qquad
  \beta_{\rm eff} = \beta\!\left(1 + \frac{\ell_1}{\ell_2}\right).
\end{equation}
Although in this case the comonotone partition function allows for a complete 
single theory description, we will use this example to demonstrate
the more general two theory description, the Schwinger--Keldysh wormhole geometry.

\paragraph{The Schwinger--Keldysh geometry.}

The gravitational geometry that computes $C_{\max}$ is built from the
Schwinger--Keldysh contour
\begin{equation}\label{eq:SK-contour}
  \underbrace{\ee^{-\beta H_1}}_{\text{Eucl.\ cap 1}}
  \;\cdot\; \underbrace{U}_{\text{Lorentzian}}
  \;\cdot\; \underbrace{\ee^{-\beta H_2}}_{\text{Eucl.\ cap 2}}
  \;\cdot\; \underbrace{U^\dagger}_{\text{Lorentzian}},
\end{equation}
where the comonotone unitary $U$ is represented as an adiabatic
time-evolution operator\footnote{In a previous version of these notes there
was a claim that $U_{\rm com}$ is implemented by adiabatic time-evolution
and not $U^{\dagger}_{\rm com}$. This was caused by the fact that after not doing any 
quantum mechanics calculations for several years I had forgotten the standard convention
that quantum amplitudes are read from right to left, not left to right.}
\begin{equation}\label{eq:U-adiabatic}
  U_{\rm com}^{\dagger} \sim \mathcal{T}\,\ee^{-i\int_0^\tau \dd t'\,
  H_{\rm int}(t')}\,,
\end{equation}
where $H_{\rm int}(t')$ is a time-dependent Hamiltonian that
interpolates between $H_1$ at $t' = 0$ and $H_2$ at $t' = \tau$.
In the adiabatic limit $\tau \to \infty$, the time-evolution
preserves the ordering of energy eigenstates: the $n$-th eigenstate
of $H_1$ is mapped smoothly to the $n$-th eigenstate of~$H_2$.
This rank-preserving property is precisely the comonotone
(Brenier) map. One subtlety in the construction is that the adiabatic time-evolution
operator is not literally the inverse of the comonotone unitary: it
carries an additional energy-dependent dynamical phase, diagonal in
the energy basis.  The phase is harmless for the BW distance: in the
combination $U\,\ee^{-\beta H_2}\,U^\dagger$ it conjugates a matrix
that the comonotone unitary has already diagonalised, and so drops
out identically.  What does distinguish the adiabatic evolution from
the comonotone unitary at finite ramp time are diabatic corrections,
which vanish as inverse powers of the ramp time
(\S\ref{sec:adiabaticity}); the adiabatic protocol is therefore
strictly suboptimal at any finite ramp time and attains the
comonotone value only asymptotically.  Both statements are made
precise at first order in perturbation theory
around~\eqref{eq:W-adiabatic-pert}.  The calculations in this section
work at the semiclassical level, where neither effect contributes;
fully quantum calculations follow in the later sections.

The interpolating Hamiltonian is largely arbitrary, provided
the adiabatic theorem holds and $H_{\rm int}$ starts at $H_1$
and ends at~$H_2$.  On the field theory side, one could construct an
interpolating Hamiltonian
$H_{\rm int}(t') =(1 - h(t'/\tau))\,H_1 + h(t'/\tau)\,H_2$ with $h(t)$ some smooth function
interpolating between 0 and 1.
On the classical gravity side, it is
sufficient that the interpolating theory admits solutions that
continuously connect the solutions of the two endpoint theories.

A concrete interpolating theory is three-dimensional gravity
coupled to a scalar:
\begin{equation}\label{eq:interpolating-action}
  S_{\rm int} = \frac{1}{16\pi G}\int \dd^3 x\,\sqrt{-g}\,
  \Bigl[\mathcal{R} - \half(\partial\varphi)^2
  - 2\,V(\varphi)\Bigr]\,,
\end{equation}
where $V(\varphi)$ interpolates smoothly between the two
cosmological constants $V_1 = -1/\ell_1^2$ and
$V_2 = -1/\ell_2^2$ over a field range~$\Delta\varphi$.
At the two endpoints $V'= 0$, so the scalar is non-dynamical
and the theory reduces at the classical level to pure gravity with the respective
cosmological constant.  The interpolation is triggered by a
time-dependent boundary condition $\varphi|_{r\to\infty} = J(t)$,
where $J(t)$ varies slowly over a timescale $\tau \gg \beta$,
driving the effective cosmological constant from $V_1$
to~$V_2$ adiabatically.

This construction is a \emph{temporal} analogue of the standard
holographic RG flow domain wall~\cite{Freedman:1999gp}.  In the
radial domain wall, the scalar interpolates between two AdS
vacua along the holographic coordinate~$r$, and each constant-$r$
slice sits at a different point along the RG flow.  Here, the
interpolation runs along the \emph{time} direction: each
constant-$t$ slice is an approximate equilibrium of the
instantaneous theory, and the adiabatic condition ensures
quasistatic evolution through the family of black hole solutions.
The breaking of conformal invariance at intermediate times
($V' \neq 0$) is physical --- the interpolation passes through a
non-conformal theory, not through a conformal manifold.

One might wonder if we are doing something inconsistent by performing an RG-flow-like operation first forwards in time (with $U^\dagger$) and then backwards (with $U$). The key difference between adiabatically interpolating between theories and performing RG flow is that RG genuinely traces over degrees of freedom, generating entropy. The adiabatic evolution is unitary: it preserves a one-to-one map between states along the flow, no states are integrated out, and no entropy is produced.

In the adiabatic approximation, the metric at each instant is the BTZ
form~\eqref{eq:btz-common} with the \emph{same} horizon radius~$r_+$,
and with $\ell$ replaced by an instantaneous effective value:
\begin{equation}\label{eq:btz-adiabatic-metric}
  \dd s^2 = -(r^2 - r_+^2)\,\dd t^2
  + \frac{\ell(t)^2}{r^2 - r_+^2}\,\dd r^2
  + r^2\,\dd\phi^2\,,
\end{equation}
with instantaneous energy $E(t) = r_+^2/(8G\,\ell(t))$.  We emphasise
that $\ell(t)$ is not a boundary condition but a \emph{dynamical}
quantity: it denotes the instantaneous effective AdS radius set by the
scalar potential, $\ell(t)^{-2} = -V(\varphi(t))$.  The boundary data
are the fixed conformal class $g^{(0)} = -\dd t^2 + \dd\phi^2$ and the
scalar source $J(t)$; the $\ell(t)^2$ that appears in~$g_{rr}$ is the
response.  The constancy of~$r_+$ is precisely the statement of
adiabaticity: the black hole adjusts quasi-statically to the changing
AdS radius without growing or shrinking.  The slow-roll corrections
and their adiabatic scaling are derived in
Appendix~\ref{app:slow-roll}.

The full piecewise metric on the SK contour is
\begin{equation}\label{eq:SK-piecewise}
\dd s^{2} =
\begin{cases}
\displaystyle
  f(r)\,\dd\tau^2 + \frac{\ell_1^2\,\dd r^2}{f(r)} + r^2\,\dd\phi^2,
  & \text{Cap 1 (Euclidean, theory 1),}
\\[8pt]
\displaystyle
  -f(r)\,\dd t^2 + \frac{\ell(t)^2\,\dd r^2}{f(r)} + r^2\,\dd\phi^2 + \mathcal{O}(\tau^{-2}),
  & \text{Lorentzian segments,}
\\[8pt]
\displaystyle
  f(r)\,\dd\tau^2 + \frac{\ell_2^2\,\dd r^2}{f(r)} + r^2\,\dd\phi^2,
  & \text{Cap 2 (Euclidean, theory 2),}
\end{cases}
\end{equation}
where $f(r) = r^2 - r_+^2$.  The
pieces are joined at corners where the Skenderis--van~Rees matching
conditions~\cite{Skenderis:2009a} --- continuity of the induced
metric and analytic continuation of the momenta --- are satisfied as long
as we start the scalar field evolution in the Lorentzian part smoothly 
from rest.  A single horizon radius $r_+$ threads
through all four segments.

\begin{figure}[t]
\centering
\includegraphics[width=0.75\textwidth]{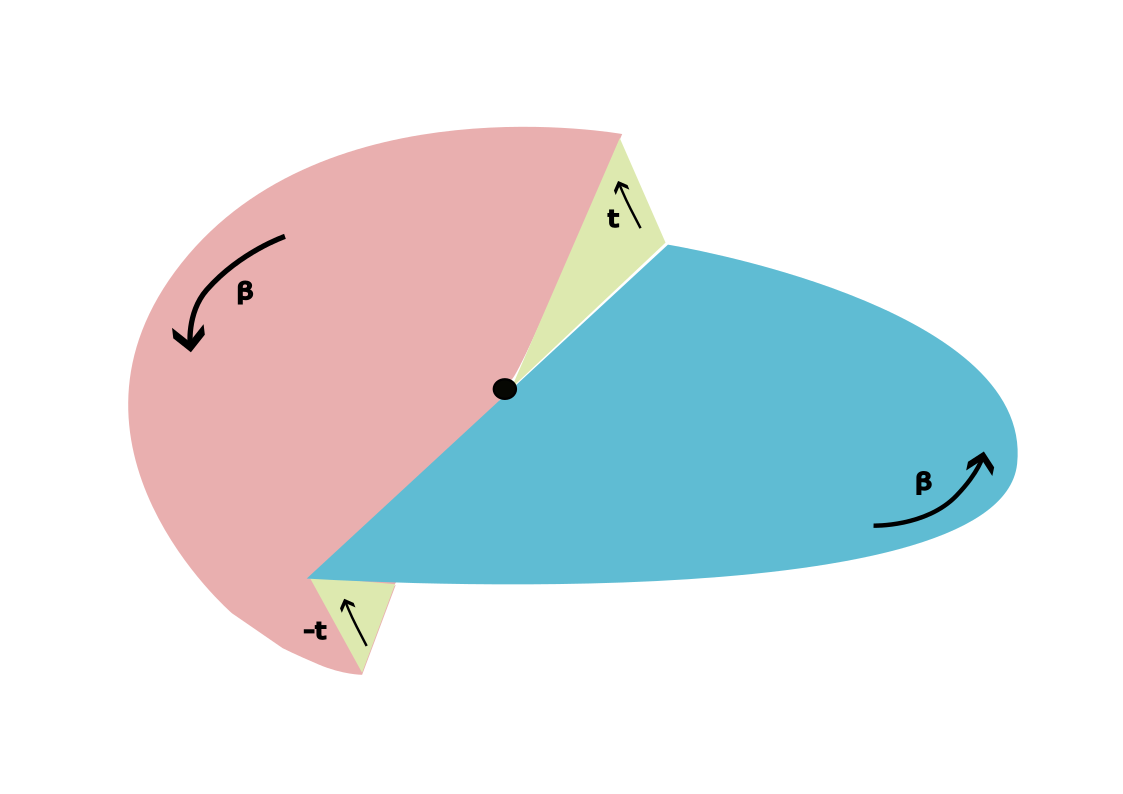}
\caption{The Schwinger--Keldysh wormhole geometry.  The blue and red
Euclidean half-discs are caps from theory~1 ($\ell_1$) and theory~2
($\ell_2$).  The green regions are Lorentzian segments where
$\ell(t)$ interpolates adiabatically.  The horizon (black dot)
stays inert during the Lorentzian evolution, which near the horizon
corresponds to a boost.}
\label{fig:sk-geometry}
\end{figure}

\paragraph{The smoothness condition.}

Expanding the metric~\eqref{eq:SK-piecewise} near the horizon
gives Euclidean angle $\theta = r_+\tau/\ell$ in each instantaneous
BTZ (or $\theta = i r_+ t/\ell$ in Lorentzian signature), so each
segment of the SK contour with Euclidean period~$\beta$ subtends
\begin{equation}\label{eq:contour-angles}
  \Delta\theta_1 = \frac{\beta r_+}{\ell_1}\,,
  \qquad
  \Delta\theta_2 = \frac{\beta r_+}{\ell_2}\,,
  \qquad
  \Delta\theta_{\rm fwd} = -\Delta\theta_{\rm bwd}
  = i\!\int_0^T\!\frac{r_+}{\ell(t)}\,\dd t\,.
\end{equation}
The Lorentzian contributions are pure imaginary (Rindler boosts) and
cancel exactly.  The total real angle wrapping the horizon is
\begin{equation}\label{eq:total-angle}
  \Theta = \Delta\theta_1 + \Delta\theta_2
  = \beta r_+\!\left(\frac{1}{\ell_1}
  + \frac{1}{\ell_2}\right)
  = 2\pi\beta(T_1 + T_2)\,,
\end{equation}
where $T_i = r_+/(2\pi\ell_i)$ is the Hawking temperature
of theory~$i$ at horizon radius~$r_+$.  Requiring smoothness
$\Theta = 2\pi$ (no conical singularity) fixes
\begin{equation}\label{eq:horizon-btz}
  r_+ = \frac{2\pi g}{\beta}\,,\qquad
  g = \frac{\ell_1\ell_2}{\ell_1 + \ell_2}\,,
\end{equation}
which is exactly the saddle-point condition~\eqref{eq:Tmc-sum}
specialised to BTZ: $T_1 + T_2 = 1/\beta$.

\paragraph{On-shell action.}

The Lorentzian on-shell actions cancel between forward and return
trips.  The Euclidean caps
contribute the standard BTZ on-shell actions, giving
\begin{equation}
  \ln C_{\max} \approx -\beta E_1 - \beta E_2
  + \frac{2\pi r_+}{4G}\,,
\end{equation}
which is exactly~\eqref{eq:C-gravity}.  The Brenier map arises from
the shared horizon: $r_+^2 = 8G\ell_1 E_1 = 8G\ell_2 E_2$,
so $E_2 = (\ell_1/\ell_2)\,E_1 = T_0(E_1)$.  It is the common
horizon that implements the Brenier map and underpins the adiabatic
construction.

The BW distance is
\begin{equation}\label{eq:W2-BTZ}
  \mathcal{W}^2_{\rm BTZ}(\beta)
  = Z_1(2\beta) + Z_2(2\beta)
  - 2\,Z_1(\beta_{\rm eff}).
\end{equation}
At high temperature the normalised distance approaches 1 (the denser
spectrum of theory 2 overwhelms theory 1), while at low temperature
it remains $O(1)$ due to the different vacuum energies --- theories
with different central charge are far apart at all temperatures.

\subsection{General adiabatic construction}
\label{sec:SK}

The BTZ example illustrates a general pattern.  The comonotone
unitary $U^{\dagger}$ admits a semiclassical dynamical representation as adiabatic
time-evolution along a family of Hamiltonians $H(\lambda)$
interpolating between $H_1$ and $H_2$.  In the adiabatic limit, the
evolution preserves the eigenstate ordering and implements the
rank-pairing.  The Schwinger--Keldysh geometry is then built from
the same four segments as~\eqref{eq:SK-contour}: two Euclidean caps
and two Lorentzian folds.

The key properties, which we saw concretely for BTZ, hold in general:
the Lorentzian segments change the boundary conditions (cosmological
constant, cutoff position, or other bulk parameters) while preserving
the horizon at fixed area $A = 4GS$.  Their on-shell actions cancel
between forward and return trips.  The smoothness condition at the
horizon, $\Theta_1 + \Theta_2 = 2\pi$, reproduces the saddle-point
equation $T_{{\rm mc},1} + T_{{\rm mc},2} = 1/\beta$
from~\eqref{eq:Tmc-sum}.  And the Brenier map arises as the energy
shift produced by changing the boundary data at fixed horizon area:
$T_0(E_1) = E(\lambda_2, S(\lambda_1, E_1))$.

\subsection{Wormhole length, decorrelation, and adiabaticity bounds}
\label{sec:wormhole-length}

The horizon of the SK wormhole is a fixed point of both Euclidean
rotations and Lorentzian boosts: near the tip of the cigar geometry,
the radial coordinate $\rho \to 0$ and the angular variable becomes
degenerate.  The extent of the Lorentzian section $\tau$ does not change the
local geometry at the throat.  Nevertheless, it produces a physical
effect: operators inserted on opposite sides of the horizon are
shifted by a boost rapidity $\eta = r_+\tau/\ell$ in Lorentzian
time, and
correlations across the horizon decay as
\begin{equation}\label{eq:decorrelation}
  \langle \mathcal{O}(A)\,\mathcal{O}(B)\rangle_{\mathrm{SK}}
  \;\sim\; \ee^{-\eta\,\Delta}\,,
\end{equation}
where $\Delta$ is the conformal dimension.  This is standard thermal
decorrelation: the SK correlator equals the analytic continuation of
the thermal two-point function to real Lorentzian time~$\eta$.

The relevant notion of distance is therefore the analytically
continued geodesic distance, not the path length through the tip
(which is trivially zero at $\rho = 0$).  In Lorentzian signature,
the infimum of path lengths between two spacelike-separated points
vanishes (attained by kinked null paths), so path length does not
control correlators.  What grows with $\eta$ is the analytically
continued geodesic distance, which for the BTZ wormhole follows
from~\eqref{eq:btz-adiabatic-metric} via
\begin{equation}\label{eq:geodesic-ac}
  \cosh(d/\ell) = u^2 + (u^2 - 1)\cosh(r_+\Delta t/\ell)\,,
\end{equation}
where $u = r/r_+$ is the dimensionless radial coordinate.  For fixed
$u > 1$ and large boost rapidity, the geodesic distance grows
linearly, $d \approx r_+\,\eta + r_+\ln\!\bigl((u^2-1)/2\bigr)$, and
the two-point function decays as $\ee^{-\eta\Delta}$.  The wormhole
``length'' in this sense is proportional to the boost rapidity: the extent of the
Lorentzian time segment controls how far apart the two boundaries are,
measured by correlations rather than by local geometry at the throat.

\paragraph{Adiabaticity bounds on $\tau$.}
\label{sec:adiabaticity}
A strict, microstate-resolved adiabatic theorem requires the
exponentially long Lorentzian time
$\tau \gg (\Delta E)^{-2} \sim \ee^{2S}$ set by the level spacing
$\Delta E \sim \ee^{-S}$; the semiclassical SK geometry cannot resolve
this scale and instead implements the Brenier map thermodynamically,
coarse-grained over energy windows $\delta E \gg \ee^{-S}$.  
At this thermodynamic level the relevant
scale is instead the lowest $s$-wave QNM frequency
$\omega_{\rm QNM} \sim 2\pi/\beta$, controlling collective
excitations: the wormhole length is bounded below by
$\tau \gtrsim 1/\omega_{\rm QNM}$ with a prefactor that grows with
the size of the spectral perturbation.  The $s$-wave QNM bound assumes the
interpolation preserves the symmetries of both theories so that
selection rules decouple the possibly gapless non-$s$-wave modes.

\subsection{Example: $T\bar{T}$ deformation}
\label{sec:ttbar}

The $T\bar{T}$ deformation~\cite{Zamolodchikov:2004ce,Smirnov:2016lqw,Cavaglia:2016oda}
is a universal irrelevant deformation of a two-dimensional CFT,
with leading-order deformed action
\begin{equation}\label{eq:ttbar-cft-action}
  S_\mu = S_{\rm CFT} + \mu\!\int\!\dd^2 x\,(T\bar T)(x) + O(\mu^2),
\end{equation}
where $(T\bar T)$ is a composite renormalized operator.  The finite-$\mu$ action is
defined recursively along the flow and is non-local in standard CFT
terms, but the deformation is solvable: the finite-volume energy
spectrum is determined exactly in terms of the undeformed energies.

On the gravity side, the holographic dual proposed
in~\cite{McGough:2016lol} is the standard AdS$_3$ Einstein--Hilbert
action with Gibbons--Hawking--York term,
\begin{equation}\label{eq:ttbar-bulk-action}
  S_{\rm bulk} = \frac{1}{16\pi G}\!\int_{r < r_c}\!\dd^3 x\sqrt{-g}\,
  \Bigl(R + \frac{2}{\ell^2}\Bigr)
  + \frac{1}{8\pi G}\!\int_{r = r_c}\!\dd^2 x\,\sqrt{-h}\,
  \Bigl(K - \frac{1}{\ell}\Bigr),
\end{equation}
but evaluated with Dirichlet boundary conditions imposed at a
finite radial cutoff $r_c \sim 1/\sqrt{\mu}$ rather than at the
asymptotic boundary $r \to \infty$.  The bulk geometry is unchanged
--- it is the same BTZ black hole as in the undeformed CFT --- only
the location and boundary conditions of the asymptotic surface
differ between the two theories.  Unlike the two-BTZ comparison of
Section~\ref{sec:btz}, both boundaries here belong to a single bulk
solution, and the deformed theory is non-conformal.  We work in the standard BTZ
parameterisation $g_{tt} = -(r^2-r_+^2)/\ell^2$, $g_{\phi\phi} = r^2$
throughout this subsection.  The energy of the dual theory is
the Brown--York quasilocal energy
\begin{equation}\label{eq:BY-energy}
  E_{\rm BY} = \frac{1}{4G\ell^2}\!\left(r_c^2
  - r_c\sqrt{r_c^2 - r_+^2}\right),
\end{equation}
evaluated at the cutoff surface, rather than the ADM energy
$E_{\rm ADM} = r_+^2/(8G\ell^2)$ measured at infinity.
Theory~1 is the undeformed CFT (boundary at $r \to \infty$,
energy~$E_{\rm ADM}$) and theory~2 is the $T\bar{T}$-deformed theory
(boundary at~$r_c$, energy~$E_{\rm BY}$).

The $T\bar{T}$ deformation maps CFT energies $E^{(0)}_n$ to
$E_n(\mu) = (1 - \sqrt{1 - 4\mu E^{(0)}_n})/(2\mu)$.
Since the deformation preserves the Hilbert space state by state,
the Brenier map is the energy map itself:
\begin{equation}\label{eq:brenier-ttbar}
  T_0(E) = \frac{1}{2\mu}\!\left(1 - \sqrt{1 - 4\mu E}\right).
\end{equation}

\paragraph{Field-theory saddle.}
We work out the saddle of the comonotone partition function
$C_{\max} = \int\!\dd E_1\,\rho_1(E_1)\,
\ee^{-\beta(E_1 + T_0(E_1))}$ in pure field-theory variables, for
later comparison with the SK on-shell action.  The undeformed CFT
on a spatial cylinder of circumference $L$ has (non-chiral,
non-rotating) Cardy entropy at high energy
\begin{equation}\label{eq:ttbar-cardy}
  S_1(E) = \sqrt{\frac{2\pi\,c\,E\,L}{3}}\,,
\end{equation}
giving the microcanonical inverse temperature
\begin{equation}\label{eq:ttbar-beta1-cardy}
  \beta_1(E) = S_1'(E) = \frac{S_1(E)}{2E}
  = \sqrt{\frac{\pi\,c\,L}{6\,E}}\,.
\end{equation}
For the $T\bar T$ map,
\begin{equation}\label{eq:ttbar-T0-derivative}
  T_0(E_1) = \frac{1 - \sqrt{1 - 4\mu E_1}}{2\mu}\,,
  \qquad T_0'(E_1) = \frac{1}{\sqrt{1 - 4\mu E_1}}\,,
\end{equation}
so $\beta_2 = \beta_1/T_0'(E_1^*) = \beta_1\,\sqrt{1 - 4\mu E_1^*}$
at the saddle (using $T_0'(E_1) = \beta_1/\beta_2$ from
differentiating the entropy-matching condition
$S_1(E_1) = S_2(T_0(E_1))$).  The general saddle
condition~\eqref{eq:saddle-condition},
$S_1'(E_1^*) = \beta(1 + T_0'(E_1^*))$, is equivalent to
\begin{equation}\label{eq:ttbar-FT-saddle}
  \frac{1}{\beta_1} + \frac{1}{\beta_2} = \frac{1}{\beta}\,,
\end{equation}
and substituting the Cardy expression~\eqref{eq:ttbar-beta1-cardy}
gives an implicit equation determining $E_1^*$ in terms of $\beta$,
$\mu$, $c$ and $L$:
\begin{equation}\label{eq:ttbar-FT-E1star}
  \sqrt{\frac{\pi\,c\,L}{6\,E_1^*}}
  \;=\; \beta\!\left(1 + \frac{1}{\sqrt{1 - 4\mu E_1^*}}\right).
\end{equation}
This is the equation that determines the saddle point energy $E_1^*$.
The on-shell value of the spectral integral is
\begin{equation}\label{eq:ttbar-FT-onshell}
  \ln C_{\max}
  \;=\; -\beta\bigl(E_1^* + T_0(E_1^*)\bigr) + S_1(E_1^*)\,,
\end{equation}
with $S_1(E_1^*)$ the Cardy entropy
from~\eqref{eq:ttbar-cardy} at the saddle.  In the following we will
show that the gravitational SK geometry reproduces the
saddle point conditions and the saddle point value exactly.

\begin{figure}[t]
\centering
\includegraphics[width=0.9\textwidth]{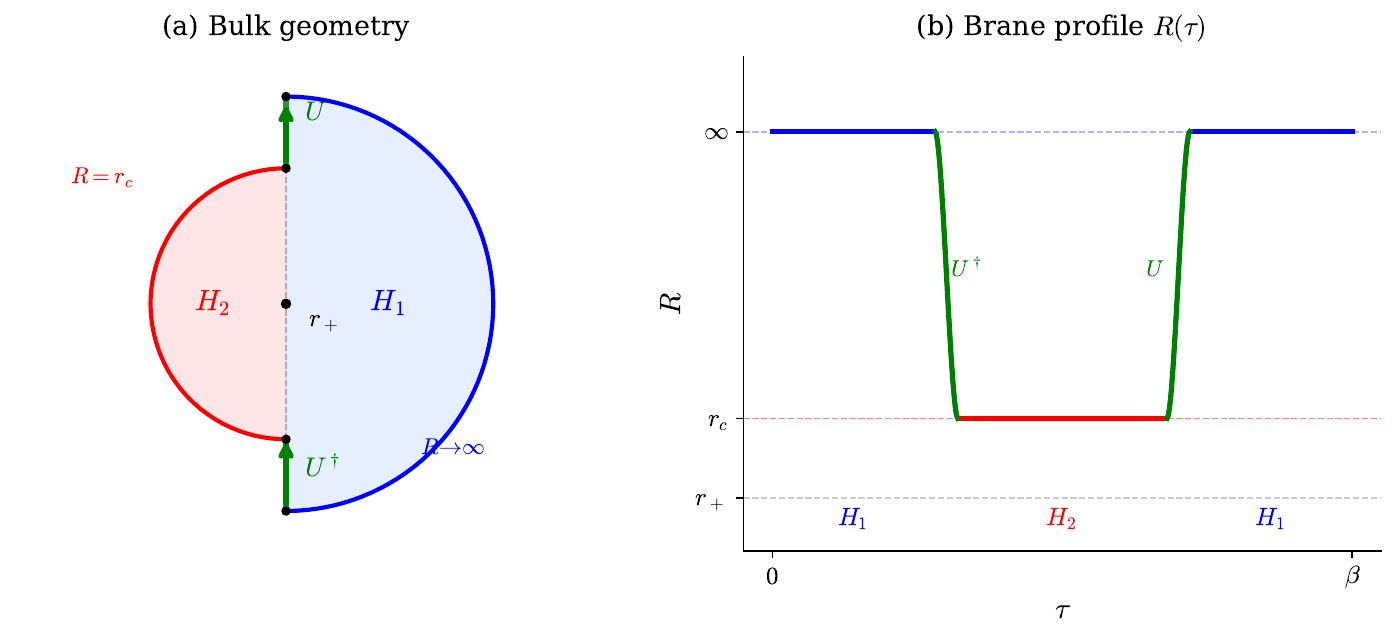}
\caption{The $T\bar{T}$ comonotone geometry.
Left: the Euclidean disc split into two caps with different
boundary radii --- Cap~1 ($H_1$, blue) at $R_{\max} \to \infty$ and
Cap~2 ($H_2$, red) at the $T\bar{T}$ cutoff $r_c$, sharing the
horizon~$r_+$.  The Lorentzian transitions (green) implement the
cutoff interpolation.
Right: the radial profile $R(\tau)$ of the cutoff surface around
the thermal circle.}
\label{fig:ttbar-brane}
\end{figure}

\paragraph{SK geometry.}
The Schwinger--Keldysh geometry has the same four-segment structure
as the BTZ case~\eqref{eq:SK-piecewise}, but with one key
difference: the blackening factor $f(r) = (r^2 - r_+^2)/\ell^2$
is the \emph{same} on every segment --- it is the radial extent of the
geometry that changes.  Cap~1 (theory~1) has its boundary at
$R_{\max} \to \infty$; Cap~2 (theory~2) has its boundary at the
$T\bar{T}$ cutoff $r_c \sim 1/\sqrt{\mu}$.  During the Lorentzian
segments, a brane at $R(t)$ interpolates between the two boundary
positions while the horizon stays at~$r_+$.  The Lorentzian on-shell
actions cancel between the outward and return trips regardless of the
brane profile~$R(t)$, provided the brane stays outside the horizon
and moves subluminally.

\paragraph{Corner matching.}
The boundary profile $R(\tau)$ and $R(t)$ is freely chosen, subject to continuity at the 
Euclidean–Lorentzian junctions and the standard SK matching condition $\dot R = 0$
at each corner~\cite{Skenderis:2009a}. With $\dot R = 0$, the boundary action density 
$\propto \sqrt{\pm\dot R^2 + f(R)^2}$ reduces to $f(R)$
on both sides of every junction, so the on-shell action receives no distributional corner contribution.

Beyond the corners, the profile $R(t)$ is subject to
two conditions: it must interpolate between $R_{\max}$ and $r_c$
with $\dot{R} = 0$ at the endpoints, and it must remain
\emph{subluminal}, $|\dot{R}(t)| < f(R(t))$, to keep the induced
metric Lorentzian.  Both conditions are easy to satisfy: choose
a smooth profile with a long enough Lorentzian time extent~$T$.
Since the asymptotic AdS region has $f(r) \sim r^2/\ell^2$, even a
null ray from $R_{\max} \to \infty$ reaches~$r_c$ in finite
coordinate time.  Since the Lorentzian actions cancel
regardless of the profile, the choice of~$T$ does not
affect~$C_{\max}$.

\paragraph{Smoothness condition.}
For a static cap with boundary at radius $R$ and asymptotic-$t$
extent $\Delta\tau$, the wedge angle subtended at the horizon
follows directly from the BTZ blackening factor $f(r) = (r^2 -
r_+^2)/\ell^2$:
\begin{equation}\label{eq:ttbar-wedge-angle}
  \theta \;=\; \frac{f'(r_+)}{2}\,\Delta\tau \;=\;
  \frac{r_+}{\ell^2}\,\Delta\tau\,.
\end{equation}
Smoothness of the SK geometry at the horizon requires the wedge
angles of the two caps to sum to a full turn,
\begin{equation}\label{eq:ttbar-smoothness-deltatau}
  \theta_1 + \theta_2 = 2\pi
  \quad\Longleftrightarrow\quad
  \frac{r_+}{\ell^2}\,(\Delta\tau_1 + \Delta\tau_2) = 2\pi\,.
\end{equation}
It remains to identify $\Delta\tau_1, \Delta\tau_2$ in terms of
the BW inverse temperature $\beta$.

\paragraph{Identifying the time extents.}
For Cap~1 the boundary is at $r \to \infty$.  The coordinate $t$ is
the natural CFT time at infinity (ADM energy is conjugate to $t$),
and the BW $\beta$ multiplies $H_1 = $ ADM mass.  Hence
$\Delta\tau_1 = \beta$ directly.

For Cap~2 the boundary is at $r = r_c$.  The induced Euclidean
metric there is
\begin{equation*}
  \dd s^2|_{r=r_c} \;=\; \frac{r_c^2 - r_+^2}{\ell^2}\,\dd\tau^2
  \;+\; r_c^2\,\dd\phi^2
  \;=\; \frac{r_c^2}{\ell^2}\!\left(\frac{r_c^2 - r_+^2}{r_c^2}\,\dd\tau^2
  + \ell^2\,\dd\phi^2\right).
\end{equation*}
We are interested in the theory whose boundary space has a fixed
length independent of the cutoff.  We therefore identify the
physical boundary metric with the Weyl-rescaled
$(\ell^2/r_c^2)\,\dd s^2|_{r=r_c}$, in which the physical Euclidean
time is
\begin{equation}\label{eq:ttbar-tauphys}
  \dd\tau_{\rm phys} \;=\; \frac{\sqrt{r_c^2 - r_+^2}}{r_c}\,\dd\tau\,,
\end{equation}
and the physical boundary spatial circle has fixed size $2\pi\ell$;
both have smooth limits as $r_c \to \infty$.  Requiring the
physical Euclidean time extent of Cap~2 to equal $\beta$ then fixes
\begin{equation}\label{eq:ttbar-deltatau2}
  \Delta\tau_2\,\frac{\sqrt{r_c^2 - r_+^2}}{r_c} \;=\; \beta
  \qquad\Longleftrightarrow\qquad
  \Delta\tau_2 \;=\; \frac{\beta\,r_c}{\sqrt{r_c^2 - r_+^2}}\,.
\end{equation}

\paragraph{Smoothness condition.}
Substituting $\Delta\tau_1 = \beta$
and~\eqref{eq:ttbar-deltatau2} into~\eqref{eq:ttbar-smoothness-deltatau}:
\begin{equation}\label{eq:ttbar-smoothness}
  \frac{\beta\,r_+}{\ell^2}
  \left(1 + \frac{r_c}{\sqrt{r_c^2 - r_+^2}}\right) = 2\pi\,.
\end{equation}
The asymmetry between the two caps reflects the redshift +
rescaling factors at the cutoff boundary.  An equivalent form,
useful for comparison with the FT saddle below, is
\begin{equation}
  \frac{1}{\beta_1} + \frac{1}{\beta_2} \;=\; \frac{1}{\beta}\,,
\end{equation}
with $\beta_1 = 2\pi\ell^2/r_+$ the standard asymptotic Hawking
inverse temperature (so $\theta_1 = 2\pi\beta/\beta_1$) and
$\beta_2 = 2\pi\ell^2\sqrt{r_c^2-r_+^2}/(r_c r_+)$ defined
analogously via $\theta_2 = 2\pi\beta/\beta_2$. 

\paragraph{On-shell action.}
The on-shell action of a BTZ black hole with a Euclidean
time-dependent finite cutoff $r = R(\tau)$ is
\begin{equation}\label{eq:action-Rtau}
  I[R(\tau)] = \frac{1}{4G}\int \dd\tau\left[
  -\frac{3R^2}{\ell^2}
  + \frac{R\,f\,\ddot{R}}{\dot{R}^2 + f^2}
  + \frac{2R^2 f^2}{\ell^2(\dot{R}^2 + f^2)}
  + \frac{R}{\ell}\sqrt{\frac{\dot{R}^2 + f^2}{f}}\right],
\end{equation}
where $f = f(R) = (R^2 - r_+^2)/\ell^2$, dots denote $\dd/\dd\tau$.

The on-shell action on the SK geometry splits into the 
two caps with Euclidean time extents $\Delta \tau_1$ and $\Delta \tau_2$
and the Lorentzian parts as
\begin{align}\label{eq:ttbar-SK-action}
I_{\rm SK} &= -\Delta \tau_1 \frac{r_+^2}{8G\ell^2}
  & &\textrm{(Cap\,1)}, \nonumber \\[4pt]
&\quad {}-\frac{i}{4G} \int \dd t\left[
  -\frac{3R^2}{\ell^2}
  - \frac{R\,f\,\ddot{R}}{-\dot{R}^2 + f^2}
  + \frac{2R^2 f^2}{\ell^2(-\dot{R}^2 + f^2)}
  + \frac{R}{\ell}\sqrt{\frac{-\dot{R}^2 + f^2}{f}}\right]
  + \textrm{c.c.}
  & &\textrm{(Lorentzian)},\nonumber \\[4pt]
&\quad {}-\Delta \tau_2 \, \frac{r_c^2 - r_c\sqrt{r_c^2 - r_+^2}}{4G\ell^2}
  & &\textrm{(Cap\,2)}.
\end{align}
The Lorentzian parts cancel exactly as long as the boundary is outside
the horizon and moves subluminally.
A short algebraic manipulation using
\eqref{eq:ttbar-smoothness} shows that
\begin{equation}\label{eq:ttbar-SK-onshell}
  I_{\rm SK} \;=\;\beta E_1 + \beta E_2 - \frac{2\pi r_+}{4G} \,,
\end{equation}
where $E_1 = \frac{r_+^2}{8G\ell^2}$ and $E_2 = \frac{r_c^2 - r_c\sqrt{r_c^2 - r_+^2}}{4G\ell^2}$.
The first two terms are the Boltzmann weight terms and the last one the Bekenstein--Hawking entropy of the
SK geometry.

\paragraph{Brenier map from the shared horizon.}
Both energies are functions of the same~$r_+$.
Eliminating~$r_+$ gives
\begin{equation}\label{eq:ttbar-brenier-gravity}
  E_2 = \frac{r_c^2}{4G\ell^2}
  \left(1 - \sqrt{1 - \frac{8G\ell^2}{r_c^2}\,E_1}\right)
  = T_0(E_1)\,,
\end{equation}
reproducing the spectral Brenier map~\eqref{eq:brenier-ttbar} upon
using the holographic dictionary $\mu = 2G\ell^2/r_c^2$.
The nonlinearity of $T_0$ arises from the square root in the
Brown--York energy at finite cutoff --- in contrast to the BTZ case,
where both energies scale as~$r_+^2$ with different prefactors,
giving a linear map.

\paragraph{Matching field theory and gravity.}
The holographic dictionary translates the gravity expressions into
the pure field-theory quantities of the previous paragraph.  The
Brown--Henneaux central charge $c = 3\ell/(2G)$ and boundary
cylinder circumference $L = 2\pi\ell$ identify the CFT data with
the asymptotic AdS geometry; the $T\bar T$ cutoff is dual to the
deformation parameter via $\mu = 2G\ell^2/r_c^2$; and the BTZ
saddle is parameterised by the horizon radius $r_+$ via the ADM
energy $E_1^* = r_+^2/(8G\ell^2)$.  Plugging these into the Cardy
expression~\eqref{eq:ttbar-beta1-cardy} gives the standard BTZ
Cardy/Bekenstein--Hawking match~\cite{Strominger:1997eq}
\begin{equation*}
  S_1(E_1^*) = \frac{\pi r_+}{2G}
\end{equation*}
i.e., the Bekenstein--Hawking entropy of the horizon.
Thus, the entropy term in gravity matches the one
in the field theory side. Finally substituting the gravitational
central charge, energy and $L$ to, the energy integral saddle
point condition \eqref{eq:ttbar-FT-saddle} we obtain exactly
the smoothness condition \eqref{eq:ttbar-smoothness}. Thus
the on-shell action of the SK geometry reproduces
the field theory saddle for $C_{\max}$.

\paragraph{Why the Lorentzian segment is essential.}
The Lorentzian part of the SK contour is not merely a convenience: it
is required to reproduce the correct comonotone partition function.  Without it,
the cutoff surface would need to move in Euclidean signature,
$R = R(\tau)$, but any $\tau$-dependent profile generates a
non-vanishing Euclidean action from~\eqref{eq:action-Rtau} that has no
counterpart in the spectral saddle-point evaluation of~$C_{\max}$.

As a concrete example, consider a piecewise constant profile
$R(\tau) = R_{\max}$ for $\tau \in [0,\Delta\tau_1]$ and
$R(\tau) = r_c$ for $\tau \in [\Delta\tau_1,\Delta\tau_1 + \Delta\tau_2]$.
Smoothing the step over an interval~$\epsilon$ and taking
$\epsilon \to 0$, the bulk action terms vanish at the junctions while
the boundary term (the $\sqrt{\dot{R}^2 + f^2}$ piece
in~\eqref{eq:action-Rtau}) produces a finite contribution
\begin{equation}\label{eq:junction-action}
  \delta I = \frac{1}{2G}\!\left(\sqrt{R_{\max}^2 - r_+^2}
  - \sqrt{r_c^2 - r_+^2}\right)
  \approx \frac{R_{\max} - r_c}{2G}\,,
\end{equation}
from the two junctions combined.
This diverges as $R_{\max} \to \infty$ and no standard counter term cancels the divergence: 
the purely Euclidean
construction is catastrophically non-optimal, producing an overlap
$C \sim C_{\max}\,\ee^{-\delta I}$ that is exponentially suppressed
relative to the true maximum.  More generally, any smooth Euclidean
profile $R(\tau)$ interpolating between $R_{\max}$ and $r_c$
contributes a positive action cost that spoils the optimality
of~$C_{\max}$.

The Lorentzian contour avoids this because the outward and return
trips cancel exactly: the brane moves from $R_{\max}$ to $r_c$ and
back, and the real Lorentzian action (guaranteed by $R(t) > r_+$)
contributes with opposite signs on the two legs.  The SK geometry
therefore realises the comonotone partition function at zero action cost for the
interpolation, with only the two Euclidean caps contributing to the
on-shell action.

\subsection{The Boltzmann--Wasserstein distance for the two examples}
\label{sec:computing-W2}

The BW distance~\eqref{eq:BW-def} is
$\mathcal{W}^2(\beta) = Z_1(2\beta) + Z_2(2\beta) - 2\,C_{\max}(\beta)$.
We work with the leading exponential of the Cardy density,
\begin{equation}\label{eq:rho-cardy-common}
  \rho_i(E) \;\approx\; \ee^{2\pi\sqrt{\ell_i\,E/(2G)}}\,.
\end{equation}
We omit the
subleading polynomial prefactor since it depends on normalisation
conventions and does not affect the qualitative features of
$\mathcal{W}^2(\beta)$ that we focus on below.  The corresponding
saddle-point partition function, to leading exponential order, is
\begin{equation}\label{eq:Z-cardy-exact}
  Z_i(\beta) \;\approx\; \ee^{\pi^2\ell_i/(2G\beta)}
\end{equation}
in the Cardy regime ($\beta \ll \pi\sqrt{\ell_i/(2G)}$), with the
saddle at $E_*^{(i)} = \pi^2\ell_i/(2G\beta^2)$.

\paragraph{BTZ.}
Since $C_{\max} = Z_1(\beta_{\rm eff})$ with
$\beta_{\rm eff} = \beta(1 + \ell_1/\ell_2)$
from~\eqref{eq:como-btz}, the distance is
\begin{equation}\label{eq:W2-BTZ-explicit}
  \mathcal{W}^2_{\rm BTZ}(\beta)
  = Z_1(2\beta) + Z_2(2\beta)
  - 2\,Z_1\!\left(\beta\!\left(1
  + \frac{\ell_1}{\ell_2}\right)\right).
\end{equation}
At high temperature, $Z_2$ dominates (it has the fastest
exponential growth $\sim \ee^{\pi^2\ell_2/(4G\beta)}$ at $\beta\to 2\beta$),
and the normalised distance $\mathcal{W}^2/(Z_1+Z_2) \to 1$: the two
theories are maximally far apart because the exponentially denser
spectrum of theory~2 cannot be matched.

\begin{figure}[t]
\centering
\includegraphics[width=\textwidth]{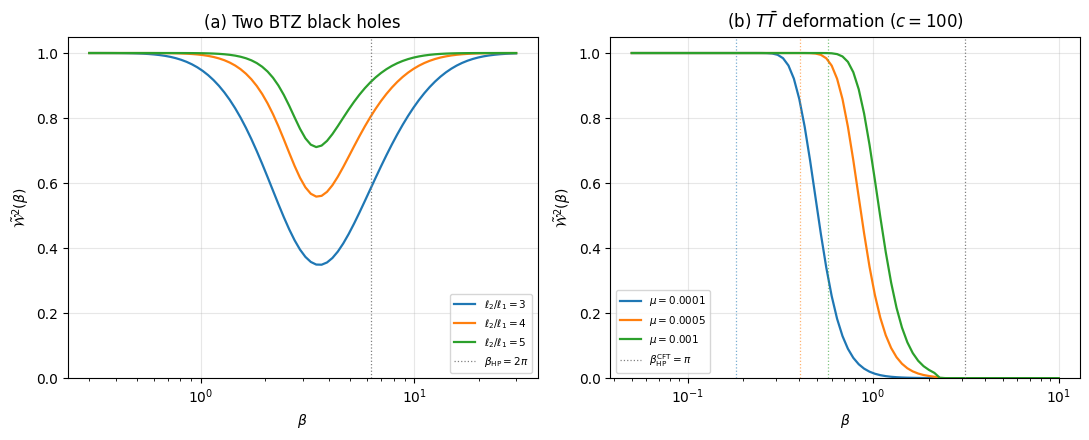}
\caption{Normalised BW distance $\tilde{\mathcal{W}}^2(\beta)$ as a
function of inverse temperature~$\beta$.
(a)~Two BTZ black holes with unequal AdS radii (including the
vacuum contribution).  The distance approaches~$1$ at both high
and low temperature; theories with different central charge are far
apart at all scales.  The minimum near the Hawking--Page transition
$\beta_{\rm HP} = 2\pi$ is deepest for nearby theories.
(b)~$T\bar{T}$ deformation.  The distance drops towards zero around
$\beta \sim (c^2\mu)^{1/3}$:
below this temperature the irrelevant deformation is invisible.}
\label{fig:W2-distance}
\end{figure}

Including the vacuum contribution
$Z_i \to \ee^{\beta\ell_i/(8G)} + Z_i^{\rm Cardy}$
(with $E_{{\rm vac},i} = -\ell_i/(8G)$), the comonotone pairs the
two vacua, adding a term
\begin{equation}\label{eq:W2-decomp}
  \mathcal{W}^2_{\rm vac}
  = \bigl(\ee^{\beta\ell_1/(8G)}
    - \ee^{\beta\ell_2/(8G)}\bigr)^{\!2},
\end{equation}
which dominates below the Hawking--Page temperature
$\beta_{\rm HP} = 2\pi$.  The normalised distance thus
approaches~$1$ at \emph{both} high and low temperature, with a
minimum near $\beta_{\rm HP}$ (Figure~\ref{fig:W2-distance}).

\paragraph{$T\bar{T}$.}
The deformed partition function and comonotone involve the
nonlinear map $T_0(E) = (1-\sqrt{1-4\mu E})/(2\mu)$ and are given
by one-dimensional integrals (evaluated numerically):
\begin{equation}\label{eq:W2-ttbar}
  \mathcal{W}^2_{T\bar{T}}(\beta)
  = Z_{\rm CFT}(2\beta) + Z_{T\bar{T}}(2\beta)
  - 2\,C_{\max}(\beta)\,.
\end{equation}
Since the $T\bar{T}$ deformation is irrelevant ($T_0(E)-E \sim
\mu E^2$), the distance vanishes exponentially at
large~$\beta$: the Boltzmann weight projects out the UV states
where the theories differ.  In particular,
$\mathcal{W}^2_{\rm vac} = 0$ because the deformation preserves
the vacuum energy.

\paragraph{$T\bar{T}$: small $\mu$ expansion.}
The saddle points of $Z_2(2\beta)$ and $C_{\max}(\beta)$
can be found and evaluated in a series expansion in $\mu$.
Writing all three quantities as integrals over the
undeformed density, $\ln Z_2 = \max_E[S(E)-2\beta T_0(E)]$ and
$\ln C_{\max} = \max_E[S(E)-\beta E-\beta T_0(E)]$, and solving the saddle
conditions order by order in $\mu$ gives, at leading exponential order,
\begin{equation}
  \ln\frac{C_{\max}(\beta)}{Z(2\beta)} = -\beta\mu E_*^2
  + \beta\mu^3 E_*^4 + O(\mu^4)\,,\qquad
  \ln\frac{Z_2(2\beta)}{Z(2\beta)} = -2\beta\mu E_*^2
  + 4\beta\mu^2 E_*^3 - 10\beta\mu^3 E_*^4 + O(\mu^4)\,.
\end{equation}
Assembling the pieces gives the BW distance,
\begin{equation}\label{eq:W2-ttbar-shifted}
  \tilde{\mathcal{W}}^2(\beta)
  = \frac{\beta^2\mu^2 E_*^4}{2}\,\bigl(1 - 4\mu E_* + O(\mu^2E_*^2)\bigr)\,.
\end{equation}
In the following sections we will use this for comparison with our later results.

\paragraph{Diagnostic of the energy scale of modification.}
The $\beta$-dependence of $\tilde{\mathcal{W}}^2(\beta) =
\mathcal{W}^2/(Z_1+Z_2)$ thus serves as a spectroscopic tool: the
temperature at which the normalised distance becomes $O(1)$ identifies
the energy scale at which the two theories begin to differ (see Figure~\ref{fig:W2-distance}).  
For $T\bar{T}$ the crossover occurs where the Boltzmann weights of the
dominant paired states differ by order one --- that is, where the energy
shift $\delta_* = T_0(E_*) - E_*$ of those states reaches the thermal
scale, $\beta\,\delta_* \sim 1$ (equivalently $\delta S(E_*)\sim 1$, cf.\
Section~\ref{sec:area-comparator}); for unequal cosmological constants
there is no crossover, as the theories differ already at the vacuum level
(although they are closest near the Hawking--Page phase transition).

\section{The area comparator}
\label{sec:area-comparator}

In the limit of small entropy
difference --- the most physically relevant case for comparing theories with 
nearby black holes --- the BW distance collapses to a strikingly simple geometric
formula: the normalised BW distance is the squared horizon-area
difference between the two black holes at equal energy.  

When the two theories have the same functional form of the density of
states but differ by a small entropy shift $\delta S(E) \ll 1$ ---
meaning $\rho_2 = \ee^{S_1 + \delta S}$ with $\delta S$ small even
though $\rho$ itself is exponentially large --- the comonotone partition function
can be evaluated directly from its defining energy integral.

The starting point is
\begin{equation}\label{eq:Cmax-energy-integral}
  C_{\max} = \int \dd E\;\ee^{-\beta(E + T_0(E))}\,\rho_1(E)\,.
\end{equation}
It is worth flagging at the outset which comparison enters where, since two
distinct ones are in play. The comonotone wormhole that defines $C_{\max}$
pairs the two theories at equal \emph{entropy} --- equal rank, equal horizon
area: the shared horizon of the Schwinger--Keldysh geometry imposes
$S_1(E_1) = S_2(E_2)$, so the paired black holes have a common area but
different energies. The $\delta A$ in the comparator is a different object
--- the area difference at equal \emph{energy}, where the two theories have
entropies differing by $\delta S = S_2(E) - S_1(E)$. What makes $\delta S$
control the result is that it enters the comonotone energy integral through
the Boltzmann weight $\ee^{-\beta T_0(E)}$, via the Brenier-shifted energy
$T_0(E) = E - T_{\rm mc}\,\delta S + \dots$ in the exponent (with $T_{\rm mc}(E) = 1/S_1'(E)$) --- not through
the shared horizon, whose area is common to both caps by construction. The
equal-entropy pairing builds the wormhole; the equal-energy area difference
reads out the distance.

Substituting the order $\delta S$ Brenier map:
\begin{equation}\label{eq:Cmax-expand}
  C_{\max} = \int \dd E\;\ee^{-2\beta E}\,\rho_1(E)\;
  \ee^{\beta\,T_{\rm mc}(E)\,\delta S(E) + O(\delta S^2)}\,.
\end{equation}
At $\delta S = 0$ this reduces to $C_{\max} = Z_1(2\beta)$: the
overlap of two identical theories, evaluated at inverse
temperature~$2\beta$. 

\begin{figure}[t]
\centering
\includegraphics[width=0.8\textwidth]{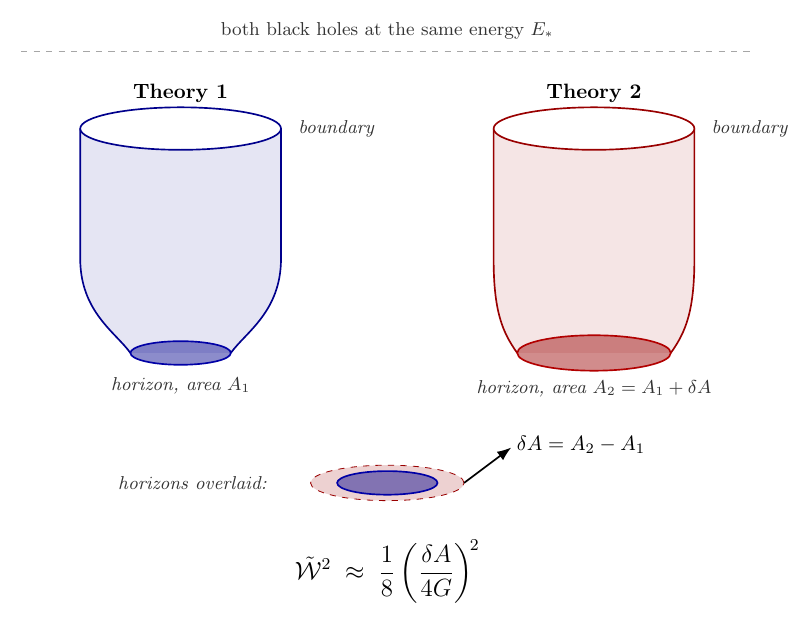}
\caption{The area comparator. The two theories are compared at the
common saddle energy $E_*$: each theory has its own Euclidean black
hole, and at equal energy their entropies differ by
$\delta S = S_2(E_*)-S_1(E_*)\ll 1$, so the horizon radii differ.
The overlaid horizons (bottom) have area difference
$\delta A = 4G\,\delta S$, and the normalised Wasserstein distance is
$\tilde{\mathcal{W}}^2 \approx \tfrac18(\delta A/4G)^2$.  These are
the horizons of two separate single-theory solutions compared at
equal energy --- not the shared horizon of the wormhole.}
\label{fig:area-comparator}
\end{figure}

For small $\delta S$, the factor
$\ee^{\beta T_{\rm mc}\delta S}$ is a slowly varying perturbation
of the integrand.  The integral is still dominated by its saddle
point at~$E = E_*$, determined by the unperturbed condition
\begin{equation}
  S_1'(E_*) = 2\beta
  \qquad\Longleftrightarrow\qquad
  2\beta\,T_{{\rm mc},1}(E_*) = 1\,.
\end{equation}
The saddle shifts by $O(\delta S)$, so to leading order we
evaluate the perturbation at~$E_*$:
\begin{equation}\label{eq:Cmax-area-expand}
  \ln C_{\max} \approx \ln Z_1(2\beta)
  + \beta\,T_{{\rm mc},1}(E_*)\,\delta S(E_*)
  = \ln Z_1(2\beta) + \tfrac{1}{2}\,\delta S\,.
\end{equation}
The factor of $1/2$ is essential:
both theories contribute symmetrically at the saddle, and the
perturbation affects only one of them.

Likewise, $Z_2(2\beta) \approx Z_1(2\beta)\,\ee^{\delta S}$
since the partition functions differ only through the entropy shift.
\footnote{Note the bookkeeping: here $\delta S$ enters through the density of
states $\rho_2 = \ee^{S_1 + \delta S}$, whereas in $C_{\max}$ above it
entered through the Boltzmann weight, via the Brenier map.  The two
placements are related by the exact change of variables
$Z_2(2\beta) = \int \dd E\,\rho_1(E)\,\ee^{-2\beta T_0(E)}$, which
moves the shift into the exponent and yields the same factor
$\ee^{\delta S}$; nothing depends on the choice.}

Assembling $\tilde{\mathcal{W}}^2 = (Z_1 + Z_2 - 2C_{\max})/ (Z_1 + Z_2)$:
\begin{equation}
  \tilde{\mathcal{W}}^2
  \approx \frac{Z_1(2\beta)(1 + \ee^{\delta S}
  - 2\,\ee^{\delta S/2})}{Z_1(2\beta) ( 1 + \ee^{\delta S})}
  =1 - \operatorname{sech}\!\left(\frac{\delta S}{2}\right)
  \approx \frac{1}{8}\,(\delta S)^2\,.
\end{equation}
Using $\delta A = 4G\,\delta S$ (the Bekenstein--Hawking relation):
\begin{equation}\label{eq:W2-area}
  {\tilde{\mathcal{W}}^2(\beta) \approx \frac{1}{8}
  \left(\frac{\delta A}{4G}\right)^{\!2},}
\end{equation}
where $\delta A$ is the difference in horizon areas \emph{of the two
theories compared at the same energy}~$E_*$.  That is: theory~1 and
theory~2 each have a black hole at energy~$E_*$, but with different
entropies $S_1(E_*)$ and $S_2(E_*) = S_1(E_*) + \delta S$, and hence
different horizon areas $A_i = 4G\,S_i(E_*)$.  The area
difference~$\delta A$ compares these two black holes, not the two caps
of the SK wormhole (which share a common horizon by construction).
The normalised Wasserstein distance is literally an area comparator (Figure~\ref{fig:area-comparator}).
\footnote{In an earlier version of these notes there was an erroneous
statement at this point, claiming that the comparison could equivalently
be made at equal temperature rather than equal energy. This is incorrect,
as one can check explicitly in the $T\bar{T}$ example.}

Keeping $\delta S$ to all orders in the exponents (still with the saddle
frozen at $E_*$)  one gets the resummed form
\begin{equation}\label{eq:W2-sech}
  \tilde{\mathcal{W}}^2 \approx 1 - \operatorname{sech}\!\left(\frac{\delta S}{2}\right),
\end{equation}
of which the quadratic comparator is the small-$\delta S$ limit.
Freezing the saddle is a good approximation whenever $\delta S$ varies
slowly on the scale of the saddle (with width
$\sigma=(-S_1'')^{-1/2}\sim E_*/\sqrt{S}$) and can be pulled out of the
spectral integral as a prefactor, i.e.\ $\delta S$ is sizeable while its
derivative $\delta S'$ is small enough not to shift the saddle point.

\paragraph{$T\bar{T}$.}
We illustrate the usage of the area comparator formula in the simple $T\bar{T}$ example.
The ``area'' of a horizon in 2+1 gravity is just $2\pi r_+$. The horizon radius $r_+$ is related to the
energy through
\begin{align}
E_{\rm BTZ} =\frac{r_+^2}{8G\ell^2},\qquad
E_{T\bar{T}} = \frac{r_c^2 - r_c\sqrt{r_c^2 - \tilde{r}_+^2}}{4G\ell^2},
\end{align}
where $r_+$ is the BTZ black hole horizon and $\tilde{r}_+$ is the $T\bar{T}$
horizon. Equating the energies $E_{\rm BTZ} = E_{T\bar{T}}$ and writing $\tilde{r}_+ = r_+ + \delta r$ and expanding in $r_+/r_c \ll 1$
(equivalently $\mu E_1 \ll 1$) we obtain,
\begin{equation}
  \delta r = -\frac{r_+^3}{8 r_c^2} + O\!\left(\frac{r_+^5}{r_c^4}\right).
\end{equation}
Using the holographic dictionary $\mu = 2G\ell^2/r_c^2$ together with
$r_+^2 = 8G\ell^2 E_1$, this is
\begin{equation}
  \delta r = -\,\tfrac{1}{2}\,r_+\mu E_1\,,
\end{equation}
so the horizon-area difference at fixed energy is
\begin{equation}\label{eq:dA-TTbar}
  \delta A = 2\pi\,\delta r = -\,\pi\, r_+\,\mu\,E_1\,.
\end{equation}
The sign is the expected one: the $T\bar{T}$ horizon sits \emph{inside}
the would-be BTZ horizon at the same energy, since some of the energy
budget is absorbed by the finite-cutoff dressing.

Plugging \eqref{eq:dA-TTbar} into the area comparator
\eqref{eq:W2-area} and using $r_+^2 = 8G\ell^2 E_1$,
\begin{equation}\label{eq:W2-area-TTbar}
  \tilde{\mathcal{W}}^2(\beta) \;\approx\;
  \frac{1}{128\,G^2}\,(\delta A)^2
  \;=\;
  \frac{\pi^2\,\ell^2\,\mu^2\,E_1^3}{16\,G}\,,
\end{equation}
We could leave the result here, but in order to compare to our earlier
result, it is convenient to remove $\ell^2/G$ with the ($2\beta$) BTZ saddle relation 
$E_1 = \pi^2\ell^2/(8G\beta^2)$ (from
$S_{\rm BTZ}'(E_*) = 2\beta$), so $\pi^2\ell^2/(8G) = \beta^2 E_*$ and
\eqref{eq:W2-area-TTbar} collapses to
\begin{equation}
  \tilde{\mathcal{W}}^2(\beta) \;\approx\;
  \frac{\beta^2\,\mu^2\,E_1^4}{2}\,,
\end{equation}
which is precisely the leading term of
\eqref{eq:W2-ttbar-shifted}.

\section{Small perturbations: spectral and operator representations}
\label{sec:perturbative}

We now consider the regime where the fractional spectral perturbation
$\delta\rho/\rho$ is small, but without assuming a semiclassical
saddle.  This covers both the tail of the semiclassical regime (many
states shifted by a small fraction) and cases where $\delta\rho$
itself is $O(1)$ (a few states in a dense spectrum).  The main
result is a direct operator representation of $\mathcal W^2$ as a
time-averaged thermal two-point function of the perturbation
(Section~\ref{sec:pert-ucom}), which serves as a fully quantum
companion to the area comparator of Section~\ref{sec:area-comparator}.
Before deriving it we set up the perturbative Brenier map and the
underlying spectral-rigidity picture, which apply in both the
many-state and few-state subregimes.

A word on scope. The algebraic statements of this section --- the diagonal
form of $X$, the spectral formula~\eqref{eq:W2-leading}, and the
time-averaged two- and four-point representations --- hold for any pair of
theories. When we \emph{evaluate} the time averages, however, we assume the
dynamics is chaotic: that connected thermal correlators decay (so their time
average is set by the conserved quantities alone), and that degeneracies do
not systematically pair states of opposite symmetry charge. In an integrable
theory the additional conserved charges contribute non-decaying pieces to
the time averages, and the corresponding statements are modified.

\subsection{Perturbative Brenier map}
\label{sec:pert-brenier}

Write $T_0(E) = E + \delta T(E)$ and expand the Brenier
condition $n_1(E) = n_2(T_0(E))$ to first order.  The right-hand side
is $n_2(E + \delta T) \approx n_1(E) + \Delta n(E) +
\rho_1(E)\,\delta T$, where
$\Delta n(E) = \int_0^E \delta\rho(E')\,\dd E'$ is the CDF
perturbation.  Solving:
\begin{equation}\label{eq:deltaT}
  T_0(E) = E - \frac{\Delta n(E)}{\rho_1(E)} + O(\delta\rho^2).
\end{equation}
The displacement is the integrated spectral density difference,
weighted by the inverse local density: where $\rho_1$ is small
(spectral edge), even a modest $\delta\rho$ produces a large
displacement; where $\rho_1$ is large (bulk of the spectrum), the map
stays close to the identity.

At first order in $\delta\rho$, the comonotone partition function
is the average of the two:
\begin{equation}\label{eq:ZBr-first-order}
  C_{\max}(\beta) = \half\,Z_1(2\beta) + \half\,Z_2(2\beta)
  + O(\delta\rho^2).
\end{equation}
To see this, substitute~\eqref{eq:deltaT} into
$C_{\max} = \int \rho_1\,\ee^{-\beta(E + T_0(E))}\,\dd E$ and
integrate by parts; the $\Delta n/\rho_1$ correction distributes
symmetrically between the two partition functions.  Since
$\mathcal{W}^2 = Z_1(2\beta) + Z_2(2\beta) - 2C_{\max}$, the
first-order term cancels exactly and the distance starts at second
order.

At first sight, equation~\eqref{eq:ZBr-first-order} is rather
striking: an expectation value in theory~1 reproduces the averaged
partition functions of both theories.  It has, however, an elementary
reading.  Under the comonotone pairing, $C_{\max}$ is the sum of
\emph{geometric} means of the paired Boltzmann weights,
\begin{equation}
  C_{\max}(\beta)
  = \sum_n \Bigl(\ee^{-2\beta E_n^{(1)}}\,
    \ee^{-2\beta E_n^{(2)}}\Bigr)^{1/2}\,,
\end{equation}
while $\half Z_1(2\beta) + \half Z_2(2\beta)$ is the sum of their
\emph{arithmetic} means; the two agree precisely to first order in
the spectral difference.

The leading Wasserstein distance can be obtained cleanly from the
perfect-square formula.  In the quantile variable $n = n_1(E)$,
\begin{equation}\label{eq:W2-perfect-square}
  \mathcal{W}^2(\beta)
  = \int_0^{N_1} \dd n\, \bigl[\ee^{-\beta E_1(n)}
  - \ee^{-\beta E_2(n)}\bigr]^2\,,
\end{equation}
which is manifestly non-negative.  Expanding
$E_2(n) = E_1(n) - \Delta n(E_1(n))/\rho_1(E_1(n))
+ O(\delta\rho^2)$, the integrand at leading order is
$[\beta\,\Delta n/\rho_1]^2\,\ee^{-2\beta E}$, giving
\begin{equation}\label{eq:W2-leading}
  {\mathcal{W}^2(\beta) = \beta^2 \int_0^\infty \dd E\,
  \frac{[\Delta n(E)]^2}{\rho_1(E)}\;\ee^{-2\beta E}\,
  + O(\delta\rho^3).}
\end{equation}
The structure is transparent: $[\Delta n(E)]^2 = [n_1(E) - n_2(E)]^2$
is the squared difference in the number of states below energy~$E$.
The factor $1/\rho_1(E)$ is the mean level spacing, which converts the
count mismatch $\Delta n(E)$ into an energy displacement
$\delta T = \Delta n/\rho_1$: a given mismatch in the number of states
costs little energy where the levels are densely packed, and much more
near the sparse spectral edge.

This formula is the fully quantum version of the semiclassical area
comparator~\eqref{eq:W2-area}.  In the semiclassical regime, $\Delta n$
can be evaluated by saddle point:
$\Delta n(E) \approx \rho_1(E)\,\delta S(E)/\beta_{\rm mc}(E)$, and
the integral localises at $\beta_{\rm mc} = 2\beta$, reproducing
$\tilde{\mathcal{W}}^2 \approx (\delta S)^2/8$. 

A concrete example of the spectral
formula~\eqref{eq:W2-leading} --- a single-microstate insertion
realised as an FZZT brane in the JT gravity matrix integral, with
the corresponding gravitational picture
is given in Appendix~\ref{app:fzzt}.

\subsection{Real-time two-point function representation of the Wasserstein distance}
\label{sec:pert-ucom}

In this subsection we consider the case where the two theories have Hamiltonians that differ by an operator
$\varepsilon V$, so that $H_2 = H_1 + \varepsilon V$. In the perturbative regime the BW distance admits an
operator representation:
\begin{equation}\label{eq:W2-time-avg-preview}
  \mathcal{W}^2 = \varepsilon^2\beta^2
  \lim_{T\to\infty}\frac{1}{T}\int_0^T \dd t\;
  \Tr\!\,\bigl(V(0)\,V(t)\,\ee^{-2\beta H_1}\bigr)
  + O(\varepsilon^3)\,,
\end{equation}
the time-averaged thermal two-point function of the perturbation
$V$ on a thermal circle of circumference~$2\beta$.
This is the perturbative analogue of the adiabatic SK construction of
Section~\ref{sec:SK}: the comonotone unitary is realised algebraically
as a small rotation in Hilbert space, and the resulting
$\mathcal W^2$ is a directly measurable thermal observable.
We derive~\eqref{eq:W2-time-avg-preview} below, then verify it on the
small-$T\bar T$ example and compare against earlier saddle point results.

Let $\{|n\rangle\}$ be the eigenstates of~$H_1$ with eigenvalues $E_n$,
and write $V_{mn} = \langle m|V|n\rangle$.  Degeneracies are handled once and for all by a choice of basis: within
each eigenspace of $H_1$ we use the basis that diagonalises the block
$P_E V P_E$ --- the adapted basis of degenerate perturbation theory ---
so that $V_{mn} = 0$ whenever $E_m = E_n$ with $m \neq n$.  As the
comonotone unitary is not unique within a degenerate subspace, we fix
the non-uniqueness by working in the adapted basis --- an explicit
choice of gauge.  With this convention all sums over $m \neq n$ below
effectively run over $E_m \neq E_n$, and no energy denominators vanish.

Standard perturbation theory then gives the eigenvalues and eigenstates
of~$H_2$ to first order:
\begin{align}
  E_n^{(2)} &= E_n + \varepsilon V_{nn} + O(\varepsilon^2)\,,
  \label{eq:En2}\\
  |n'\rangle &= |n\rangle + \varepsilon \sum_{m\neq n}
    \frac{V_{mn}}{E_n - E_m}\,|m\rangle + O(\varepsilon^2)\,.
  \label{eq:n-prime}
\end{align}
The comonotone unitary maps
ordered eigenstates of~$H_2$ to ordered eigenstates of~$H_1$:
$U_{\mathrm{com}}|n'\rangle = |n\rangle$.  Since the ordering is
preserved by continuity for small~$\varepsilon$, this gives
\begin{equation}\label{eq:Ucom-pert}
  U_{\mathrm{com}} = \mathbf{1} - i\varepsilon A + O(\varepsilon^2)\,,
  \qquad
  A_{mn} = -\,\frac{i\,V_{mn}}{E_n - E_m}\;\;(E_m \neq E_n)\,,
  \quad A_{mn} = 0 \text{ otherwise}\,.
\end{equation}
The generator~$A$ is Hermitian (so $U_{\mathrm{com}}$ is unitary
through~$O(\varepsilon)$) and is precisely the adiabatic gauge
connection --- the Berry connection on the space of theories,
evaluated at $H_1$.

This makes the relation to the adiabatic construction of
Section~\ref{sec:SK} precise at first order.  Consider the
interpolating Hamiltonian $H(t) = H_1 + \varepsilon(t)\,V$ with a slow
ramp satisfying $\varepsilon(0)=0$, $\varepsilon(\tau)=\varepsilon$.
At first order in time-dependent perturbation theory, the
off-diagonal matrix elements of the evolution operator are controlled
by the Fourier weight of the ramp at the transition frequency,
$\hat\varepsilon(\omega) = \int_0^\tau \dd t\,\varepsilon(t)\,
\ee^{-i\omega t}$.  Adiabaticity fixes this weight universally:
integrating by parts, the interior of the ramp averages out and only
the endpoint $t=\tau$ survives,
\begin{equation}\label{eq:ramp-weight}
  \hat\varepsilon(\omega)
  = \frac{i\,\varepsilon}{\omega}\,\ee^{-i\omega\tau}
  \Bigl[\,1 + O\bigl((\omega\tau)^{-n}\bigr)\Bigr]\,,
\end{equation}
with $n = k+1$ for a ramp whose first $k$ derivatives vanish at the
endpoints ($n=1$ for a linear ramp): each integration by parts trades
a ramp derivative, of size $1/\tau$, for a factor $1/\omega$.  Note
that the magnitude $\varepsilon/\omega$ is independent of how slowly
the ramp is taken --- slowness only suppresses the corrections and
feeds the never-decaying phase.  Using~\eqref{eq:ramp-weight},
\begin{equation}\label{eq:W-adiabatic-pert}
  \bigl\langle m\bigl|\,\mathcal{T}\ee^{-i\int_0^\tau \dd t\,H(t)}
  \bigr|n\bigr\rangle
  = \Bigl(\delta_{mn}
  + \frac{\varepsilon\,V_{mn}}{E_n - E_m}\Bigr)\,\ee^{-i\theta_n}
  + O(\varepsilon^2)\,,
  \qquad
  \theta_n = E_n\tau
  + V_{nn}\!\int_0^\tau\!\varepsilon(t)\,\dd t\,.
\end{equation}
The bracket is exactly $(U_{\mathrm{com}}^\dagger)_{mn}$
from~\eqref{eq:Ucom-pert}: the adiabatic time-evolution operator
equals $U_{\mathrm{com}}^\dagger D$ with
$D = \mathrm{diag}(\ee^{-i\theta_n})$ collecting the dynamical
phases.  Since $D$ is diagonal in the energy basis, it drops out of
the combination $U\,\ee^{-\beta H_2}\,U^\dagger$ identically, and the
adiabatic evolution computes the same BW distance as the comonotone
unitary, up to diabatic corrections vanishing as inverse powers
of~$\tau$.

Define $X = \ee^{-\beta H_1} - U_{\mathrm{com}}\,\ee^{-\beta H_2}\,
U_{\mathrm{com}}^\dagger$.  At $O(\varepsilon)$ there are two
contributions to its matrix elements $X_{mn}$.

The first comes from the perturbed spectrum (setting $U = \mathbf{1}$
but keeping the full $H_2$).  By first-order perturbation theory the
matrix $\ee^{-\beta H_2}$ in the $|n\rangle$ basis has off-diagonal
elements
$\langle m|\ee^{-\beta H_2}|n\rangle = \varepsilon\,V_{mn}({\ee^{-\beta
E_n} - \ee^{-\beta E_m}})/({E_n - E_m})$ for $E_m \neq E_n$ (the
equal-energy off-diagonal elements vanish by the adapted-basis
convention), and diagonal
shift $\langle n|\ee^{-\beta H_2}|n\rangle - \ee^{-\beta E_n} =
-\varepsilon\beta V_{nn}\ee^{-\beta E_n}$.

The second comes from the unitary rotation (keeping $\ee^{-\beta H_1}$
but including $U_{\mathrm{com}}$):
$U\,\ee^{-\beta H_1}\,U^\dagger = \ee^{-\beta H_1} -
i\varepsilon[A,\,\ee^{-\beta H_1}] + O(\varepsilon^2)$.  With $A_{mn}$
from~\eqref{eq:Ucom-pert}, the off-diagonal of
$-i\varepsilon[A,\ee^{-\beta H_1}]$ evaluates to
$-\varepsilon\,V_{mn}({\ee^{-\beta E_n} - \ee^{-\beta E_m}})/
({E_n - E_m})$, again with vanishing equal-energy elements.

The two off-diagonal contributions to $-U\ee^{-\beta H_2}U^\dagger$ ---
which carry an overall minus sign in $X$ relative to the matrix elements
listed above --- have opposite signs and \emph{cancel exactly}.  Only
the diagonal survives:
\begin{equation}\label{eq:X-diag}
  X_{mn} = \delta_{mn}\,\varepsilon\,\beta\,V_{nn}\,\ee^{-\beta E_n}
  + O(\varepsilon^2)\,.
\end{equation}
The physical content is clear: the comonotone unitary undoes the basis
rotation caused by the perturbation, isolating the genuine eigenvalue
shifts $\delta E_n = \varepsilon V_{nn}$ --- for a degenerate level,
the eigenvalues of the block $P_E V P_E$, exactly as in degenerate
perturbation theory.  Since $X$ is diagonal,
\begin{equation}\label{eq:W2-pert-discrete}
  \mathcal{W}^2 = \Tr(X^2) = \beta^2 \sum_n
  (\delta E_n)^2\,\ee^{-2\beta E_n}\,.
\end{equation}
In the continuum limit this becomes~\eqref{eq:W2-leading}: each eigenvalue near energy~$E$
shifts by $\delta E$ leading to a decrease in the number of states below $E$ 
given by $\Delta n(E)= -\delta E \rho(E)$, and the sum over
$\rho(E)\,\dd E$ levels per interval reproduces the
$[\Delta n]^2/\rho$ integrand.

The formula~\eqref{eq:W2-pert-discrete} involves only the diagonal
matrix elements $V_{nn}$.  It can be recast as a time-averaged thermal
two-point function:
\begin{equation}\label{eq:W2-time-avg}
  \mathcal{W}^2 = \varepsilon^2\beta^2
  \lim_{T\to\infty}\frac{1}{T}\int_0^T \dd t\;
  \Tr\,\!\bigl(V(0)\;V(t)\;\ee^{-2\beta H_1}\bigr)\,,
\end{equation}
where $V(t) = \ee^{iH_1 t}\,V\,\ee^{-iH_1 t}$.  To see that this
reproduces~\eqref{eq:W2-pert-discrete}, insert a complete set: the
integrand becomes
$\sum_{m,n}|V_{mn}|^2\,\ee^{i(E_n - E_m)t}\,\ee^{-2\beta E_m}$, and
the $1/T$ time average kills all phases with $E_m \neq E_n$, while the
surviving equal-energy off-diagonal terms vanish by our choice of
basis, leaving only the diagonal $V_{nn}^2$ terms.

Now recognise the structure of~\eqref{eq:W2-time-avg}: the factor
$\ee^{-2\beta H_1}$ is a Euclidean path integral over a thermal circle
of circumference~$2\beta$ (two Euclidean caps of length~$\beta$), the
operator $V(t)$ inserts a perturbation on a Lorentzian section of
duration~$t$, and the $\int \dd t/T$ averages over all Lorentzian
durations.  This is reminiscent of the Schwinger--Keldysh contour of
Section~\ref{sec:SK}, with the difference that both caps are in
theory~1 and the perturbation enters through insertions of~$V$.
Note that the placement of the points in the two-point function inside the time average
is not unique. Many different placements of the operators on the Schwinger--Keldysh
contour can give rise to the same time average. 
What is essential is that they are separated by a Lorentzian time $t$.

\subsection{Second-order: perturbations with vanishing one-point function}
\label{sec:pert-otoc}

We now assume that the thermal one-point function of the perturbation
vanishes at the order we work, $\langle V\rangle_{2\beta}=0$. The cleanest
realisation is a symmetry: if $V$ is odd under an unbroken discrete global symmetry of
theory~1, the diagonal elements $\langle n|V|n\rangle$ vanish state by
state and the suppression is exact. (Degeneracies are harmless here:
descendants inherit the parity of their primary, so Virasoro multiplets do
not mix sectors, and accidental cross-sector degeneracies are
non-generic.) For a primary without a protecting symmetry the vanishing is
volume-dependent: it is exact in infinite volume, where the thermal line
is conformal to the plane, while at finite volume
$\langle V\rangle_{2\beta}$ is a torus one-point function and vanishes
only at leading semiclassical order, with exponentially small corrections.
Granting the assumption, the disconnected piece of the two-point function
vanishes and the $O(\varepsilon^2)$ result~\eqref{eq:W2-time-avg} reduces
to the time average of the \emph{connected} thermal correlator, which
decays exponentially on the QNM scale, so its time average vanishes as
well (up to the non-perturbative $\ee^{-S}$ accuracy). The leading non-trivial BW distance is therefore
$O(\varepsilon^4)$, where it
admits a real-time representation as a time-averaged thermal
four-point function of~$V$ with out-of-time-ordered operator
kinematics:
\begin{equation}\label{eq:W2-otoc}
  \mathcal{W}^2 = -\varepsilon^4\beta^2
  \lim_{T\to\infty}\frac{1}{T}\!\int_0^T\!\!\dd s\!\int_0^\infty\!\!\dd t_1\!\int_0^\infty\!\!\dd t_2\;
  \ee^{-\eta(t_1+t_2)}\,
  \Tr\!\,\bigl(V(0)\,V(-t_1)\,V(s)\,V(s-t_2)\,\ee^{-2\beta H}\bigr)
  + O(\varepsilon^5)\,,
\end{equation}
with $V(t) = \ee^{iHt}V\ee^{-iHt}$ and $\eta\to 0^+$ a convergence
regulator.  The derivation, together with the all-orders identification
$X_{nn} = \ee^{-\beta E_n}\bigl(1 - \ee^{-\beta\delta E_n}\bigr)$
that underlies it, is given in Appendix~\ref{app:4pt}.

\subsection{Example: small $T\bar{T}$ deformation from the two-point function}
\label{sec:ttbar-pert}

As a consistency check, we compute the BW distance for a small
$T\bar{T}$ deformation using the two-point function
formula~\eqref{eq:W2-time-avg}, and verify that it reproduces the area
comparator~\eqref{eq:W2-area}.

The $T\bar T$ flow $\partial_\mu S = \int \dd^2x\,
\det T_{\mu\nu}$ --- which defines the deformation parameter $\mu$
appearing in the spectrum formula~\eqref{eq:brenier-ttbar} --- is
generated by the Hamiltonian perturbation
\begin{equation}\label{eq:ttbar-V-def}
  V = \int_0^1 \dd x\;\det T_{\mu\nu}(x)\,.
\end{equation}
For a 2d CFT on flat space the trace vanishes, $T^\mu{}_\mu = 0$, so
$T_{tt} = T_{xx}$ and
$\det T = T_{tt}T_{xx} - T_{tx}^2 = T_{tt}^2 - T_{tx}^2$.  The
thermal one-point functions are
\begin{equation}\label{eq:T-one-pt}
  \langle T_{tt}\rangle_{2\beta}
  = \langle T_{xx}\rangle_{2\beta}
  = E^*_{\rm thermal}\,,\qquad
  \langle T_{tx}\rangle_{2\beta} = 0\,,
\end{equation}
where $E^*_{\rm thermal} = \pi c/(24\beta^2)$ is the thermal energy on
the unit cylinder at inverse temperature~$2\beta$, equal to the Cardy
saddle energy of $Z(2\beta)$.  The disconnected one-point of $\det T$
is therefore
\begin{equation}\label{eq:detT-disc}
  \langle\det T\rangle_{2\beta}^{\rm disc}
  = \langle T_{tt}\rangle_{2\beta}\langle T_{xx}\rangle_{2\beta}
  - \langle T_{tx}\rangle_{2\beta}^2
  = (E^*_{\rm thermal})^2\,.
\end{equation}

We now apply the time-averaging formula~\eqref{eq:W2-time-avg}.  The
two-point function $\langle\det T(0)\,\det T(\Delta x,t)\rangle_{2\beta}$
splits into a disconnected piece and connected pieces involving
stress-tensor correlators.  

The connected correlators on the thermal cylinder decay exponentially as
$\sim \ee^{-2\pi|t|/\beta}$ (quasinormal mode decay at the gap
$\Delta = 2$ of the stress tensor), with one exception: after spatial
integration the stress tensor contains the conserved charges
($\int\!\dd x\,T_{tt} = H$, $\int\!\dd x\,T_{tx} = P$), whose connected
correlators do not decay and survive the time average.
For a conserved charge $Q(t) = Q(0)$, since the time-evolution operators
can be commuted past $Q$; the correlator is then time-independent, the
time average acts trivially, and what survives is the static thermal
variance of $Q$.
These contributions are suppressed by
one power of the central charge relative to the disconnected
contribution, and we do not attempt to match them here.  At the leading
large-$c$ order at which we work, only the fully disconnected piece
survives:
\begin{equation}\label{eq:ttbar-disc}
  \lim_{T\to\infty}\frac{1}{T}\int_0^T \dd t\int \dd(\Delta x)\;
  \langle \det T(0)\,\det T(\Delta x,t)\rangle_{2\beta}
  = (E^*_{\rm thermal})^4\,.
\end{equation}
Substituting into~\eqref{eq:W2-time-avg} with $\varepsilon = \mu$ and
using $Z_1 + Z_2 \approx 2Z(2\beta)$ to leading order in~$\mu$ (so
that $Z(2\beta)$ cancels):
\begin{equation}\label{eq:ttbar-pert-result}
  \tilde{\mathcal{W}}^2
  \;=\;
  \frac{\mathcal{W}^2}{Z_1(2\beta)+Z_2(2\beta)}
  \;=\;
  \frac{\mu^2\beta^2}{2}\,\bigl(E^*_{\rm thermal}\bigr)^4\,.
\end{equation}
This agrees with the results from the saddle point/SK geometry
analysis~\eqref{eq:W2-ttbar-shifted} and the area comparator.


\section{Discussion}
\label{sec:discussion}

We have introduced the Boltzmann--Wasserstein distance, a
temperature-resolved metric on the space of quantum theories built entirely
from partition-function data, and computed it in three ways that agree: as
an optimisation over wormholes, whose optimum is the comonotone partition
function $C_{\max}$; as a semiclassical area comparator
$\tilde{\mathcal W}^2 \approx (\delta A/4G)^2/8$, sensitive to the
rearrangement of the spectrum; and as a time-averaged correlation function
of the perturbing operator, sensitive also to its eigenvectors. The
construction is quantitative: the on-shell action of the Schwinger--Keldysh
wormhole reproduces the spectral saddle of $C_{\max}$, conditions and value
alike.

Our worked examples are in $2+1$ dimensions, but little in the construction
depends on this: the area comparator assumed only that entropy is horizon
area, and we expect higher-dimensional examples to go through with no new
obstacles. The genuinely different regime is JT gravity and its
deformations, which we have deliberately set aside (beyond
Appendix~\ref{app:fzzt}): there is typically no classical saddle that
approximates the calculation, and one must work directly at the quantum
level. The distance remains well defined there, but the details differ
enough that we leave them for separate treatment.

The BW distance as defined rests on two ingredients --- a global thermal
trace, and a strict rank-by-rank pairing of the two spectra --- and the two
extensions we find most promising each relax one of them.

Relaxing the first, one can replace $H$ by a modular Hamiltonian, defining a
subregion BW distance. This would compare entanglement wedges rather than
full theories, and in particular would allow the Wasserstein distance to be
studied within a single theory --- between states, or between subregions of
a single state.

Relaxing the second, the entropic regularisation of optimal
transport~\cite{Cuturi:2013} softens the strict monotone pairing.
Gravitationally this should interpolate between the optimal wormhole and the
Haar-random disconnected geometry --- a one-parameter family between
$C_{\max}$ and $Z_1 Z_2/N$. In a matrix model the regularised optimisation
is implemented by the Harish-Chandra--Itzykson--Zuber integral, producing a
coupled two-matrix model that is well defined and interesting in its own
right. Where we have got stuck is the double-scaling limit relevant to
gravity, which is not as straightforward as for a single matrix model; we
flag this as the concrete open problem on this route.

Finally, the adiabatic construction invites a complexity-theoretic reading.
The semiclassical wormhole realises the comonotone unitary by a slow ramp,
whose microstate-exact version requires a Lorentzian segment of length
$\tau \sim \ee^{2S}$ --- the inverse-gap-squared cost of resolving an
$\ee^{-S}$ level spacing. This is the brute-force implementation: the
rank-pairing itself is trivial in the energy eigenbasis, so the
exponential length reflects the cost of the adiabatic \emph{algorithm},
not of $U_{\rm com}$. Whether a shorter wormhole can implement the same
optimal pairing --- a counterdiabatic ``shortcut'' trading length for a
more intricate bulk source --- and whether its length is bounded below by
an intrinsic complexity of the pair of theories, we leave as an open
question.

\newpage
\appendix

\section{Slow-roll construction of the Lorentzian segment}
\label{app:slow-roll}

In this appendix we construct the Lorentzian segment of the
Schwinger--Keldysh wormhole explicitly, using a scalar field that
rolls along a potential connecting the two AdS vacua.  We work in
2+1 dimensions throughout; the extension to higher dimensions is
straightforward modulo the graviton sector discussed in
\S\ref{sec:adiabaticity}.

The construction provides three things: (i) a concrete, diffeomorphism-invariant realisation of
the adiabatic interpolation that makes no gauge choices beyond the
null coordinate; (ii) analytical control over the corrections to all
fields at each order in the slow-roll parameter~$\varepsilon$ and the
adiabatic parameter~$1/\tau$; and (iii) numerical confirmation that
the corrections vanish as $\tau\to\infty$.

\paragraph{Frame remark.}
Throughout this appendix we work in the natural BTZ parameterisation
$f_0 = (r^2-r_+^2)/\ell^2$, $g_{rr} = \ell^2/(r^2-r_+^2)$, $g_{\phi\phi}
= r^2$, because the canonical form $g_{tt}g_{rr} = -1$ makes the Vaidya
transition $v = t + r_*$ with $\dd r_*/\dd r = 1/f$ clean.  This is the
same physical bulk solution as the parameterisation of
Section~\ref{sec:btz} ($g_{tt} = -(r^2-r_+^2)$, $g_{rr} =
\ell^2/(r^2-r_+^2)$), related by the time rescaling $t \to t/\ell$.  
In particular, the proper boundary
circumference $2\pi\ell(t)$ that appears here under the slow roll is a
frame artifact: in the Section~\ref{sec:btz} frame the boundary
cylinder is fixed at proper circumference~$2\pi$ and the time
dependence sits entirely in the bulk scalar profile and the response
$\ell(t)$.  The bulk corrections derived below ($\Phi_{\rm s}$,
$\delta_{\rm s}$, $f_{\rm s}$, $\psi$, $\Delta\delta$, $\Delta f$) are
gauge-invariant statements about the bulk geometry and apply unchanged
in either frame; only their projection onto the boundary differs.

\subsection{Setup}

Consider three-dimensional gravity coupled to a scalar:
\begin{equation}\label{eq:app-action}
  S = \frac{1}{16\pi G}\int \dd^3 x\,\sqrt{-g}\,
  \bigl[R - \half(\partial\phi)^2 - 2V(\phi)\bigr]\,,
\end{equation}
with potential $V(\phi)$ interpolating between $V_1 = -1/\ell_1^2$
and $V_2 = -1/\ell_2^2$ over a field range $\Delta\phi$.  The
slow-roll parameter $\varepsilon \equiv V'(\phi)$ is bounded by
$\varepsilon_{\max} \sim |\Delta V|/\Delta\phi$.  The scalar equation
of motion is $\Box\phi = 2V'(\phi)$.

The boundary value $J(v) = \phi(v,R)$ is driven from $J_0$ (in the
$\ell_1$ vacuum) to $J_f$ (in the $\ell_2$ vacuum) over a Lorentzian
time~$\tau$, implementing the interpolation~\eqref{eq:SK-contour}.

\subsection{Formulation}

We use ingoing Eddington--Finkelstein--Vaidya coordinates, in which
the metric is regular at the horizon:
\begin{equation}\label{eq:app-vaidya}
  \dd s^2 = -f\,\ee^{-2\delta}\,\dd v^2
  + 2\,\ee^{-\delta}\,\dd v\,\dd r
  + r^2\,\dd\varphi^2\,.
\end{equation}
The Einstein equations decompose into two radial constraints, solved
on each $v$-slice for given scalar profile $\Phi(v,r)$:
\begin{align}
  \partial_r\delta &= -\frac{r}{2}\,(\partial_r\Phi)^2\,,
  \label{eq:app-C1}\\
  \partial_r f + \frac{r}{2}\,(\partial_r\Phi)^2\,f
  &= -2r\,V(\Phi)\,,\label{eq:app-C2}
\end{align}
with boundary conditions $\delta(v,R) = 0$ and $f(v,r_+) = 0$.  The
scalar wave equation $\Box\Phi = 2V'$ becomes, after eliminating
$\delta$ via~\eqref{eq:app-C1}:
\begin{equation}\label{eq:app-scalar}
  \ee^{\delta}\!\left(2\,\partial_v\partial_r\Phi
  + \frac{\partial_v\Phi}{r}\right)
  + \frac{1}{r}\,\partial_r(r f\,\partial_r\Phi)
  + \frac{r}{2}\,f\,(\partial_r\Phi)^3 = 2V'(\Phi)\,.
\end{equation}
The cubic term $\frac{r}{2}f(\partial_r\Phi)^3$ arises from the
$\delta$-dependence of the d'Alembertian via~\eqref{eq:app-C1}; it
is $O(\varepsilon^3)$ perturbatively but is required for exact
consistency of the constraint--evolution system.

Evolution proceeds by characteristic integration: defining
$W = \partial_v\Phi$, the scalar equation reduces to a first-order
radial ODE for~$W$,
\begin{equation}\label{eq:app-W}
  2\,\partial_r W + \frac{W}{r}
  = \ee^{-\delta}\!\left[2V'
  - \frac{1}{r}\,\partial_r(rf\,\partial_r\Phi)
  - \frac{r}{2}\,f\,(\partial_r\Phi)^3\right],
\end{equation}
with $W(v,R) = \dot J(v)$.  At each step, the constraints determine
$\delta$ and $f$; the $W$-equation determines $\partial_v\Phi$;
then $\Phi$ is advanced in~$v$.

\subsection{Instantaneous static solution}

At zeroth adiabatic order ($\dot J = 0$), the solution at each
instant is the static configuration $\Phi_{\rm s}(r; J)$, $f_{\rm
s}(r; J)$, $\delta_{\rm s}(r; J)$ solving the coupled
system~\eqref{eq:app-C1}--\eqref{eq:app-scalar} with $\partial_v = 0$.
In the slow-roll expansion:

\paragraph{Scalar.}
The static scalar equation reduces to
$\frac{1}{r}\partial_r(r f_0\,\Phi_{\rm s}') = 2\varepsilon$,
where $f_0 = (r^2-r_+^2)/\ell^2$ and $\varepsilon = V'(\Phi)$.
Regularity at $r = r_+$ and the boundary condition $\Phi_{\rm s}(R) = J$
give:
\begin{equation}\label{eq:app-static-phi}
  \Phi_{\rm s}(r) = J - \varepsilon\,\ell^2\,\ln\frac{R}{r}
  + O(\varepsilon^2)\,.
\end{equation}
The derivative $\Phi_{\rm s}' = \varepsilon\ell^2/r$ is independent of
the black hole mass, with a logarithmic running characteristic of a
marginal deformation.

\paragraph{Lapse.}
From~\eqref{eq:app-C1}: $\delta_{\rm s}'
= -(r/2)(\Phi_{\rm s}')^2 = -\varepsilon^2\ell^4/(2r)$, giving
\begin{equation}\label{eq:app-static-delta}
  \delta_{\rm s}(r)
  = -\frac{\varepsilon^2\ell^4}{2}\,\ln\frac{r}{R}
  + O(\varepsilon^3)\,.
\end{equation}

\paragraph{Blackening function.}
The constraint~\eqref{eq:app-C2} receives $O(\varepsilon^2)$
contributions from both the kinetic term $(r/2)(\Phi_{\rm s}')^2 f$
and the radial variation of $V(\Phi_{\rm s}(r))$.  Together they give
\begin{equation}\label{eq:app-static-f}
  f_{\rm s}(r) = \frac{r^2 - r_+^2}{\ell^2}
  - \varepsilon^2\ell^2\!\left[\left(r^2 - \frac{r_+^2}{2}\right)
  \ln\frac{r}{r_+} - \frac{r^2-r_+^2}{4}\right]
  + O(\varepsilon^3)\,.
\end{equation}
The $O(\varepsilon^2)$ correction describes a running effective
cosmological constant: $1/\ell_{\rm eff}^2(r) = 1/\ell^2 - \varepsilon^2
\ell^2\ln(r/r_+)$.

At each endpoint ($\varepsilon = 0$), the solution reduces to exact
BTZ with the appropriate $\ell_i$.  The interpolation is therefore
exact at the start and end, with $O(\varepsilon^2)$ transient
corrections during the roll.

\subsection{Leading adiabatic corrections}

We now expand around the quasi-static solution:
$\Phi = \Phi_{\rm s}(r; J(v)) + \psi(v,r)$, with $\psi(v,R) = 0$.

\paragraph{Scalar correction at $O(\dot J)$.}
Substituting into the scalar equation and using $S[\Phi_{\rm s}] = 0$,
the leading correction satisfies
\begin{equation}\label{eq:app-psi-eqn}
  \frac{1}{r}\,\partial_r(r f_{\rm s}\,\partial_r\psi) = -\frac{\dot J}{r}\,,
\end{equation}
sourced by the uniform sweep rate.  The source $\dot J/r$ arises because the
time derivative $\partial_v\Phi = \dot J$ is spatially uniform while the
W equation~\eqref{eq:app-W} demands a radial profile.  Integrating with regularity at $r_+$:
\begin{equation}\label{eq:app-psi}
  {\psi(r) = -\frac{\dot J\,\ell^2}{r_+}\,
  \ln\frac{r\,(R + r_+)}{R\,(r + r_+)}\,.}
\end{equation}
At the horizon, $\psi(r_+) \approx (\dot J\,\ell^2/r_+)\ln 2$ for
$R \gg r_+$: the scalar lags behind its equilibrium value by an
amount proportional to the sweep rate and the squared AdS radius.

\paragraph{Lapse correction at $O(\varepsilon\dot J)$.}
The cross term between $\Phi_{\rm s}'$ and $\psi'$
in~\eqref{eq:app-C1} gives
\begin{equation}\label{eq:app-delta-corr}
  \Delta\delta(r) = \frac{\varepsilon\,\dot J\,\ell^4}{r_+}\,
  \ln\frac{r\,(R+r_+)}{R\,(r+r_+)}\,.
\end{equation}
This is a mixed slow-roll${}\times{}$adiabatic correction to the
off-diagonal metric component $g_{vr} = \ee^{-\delta}$.

\paragraph{Blackening function at $O(\varepsilon\dot J)$ and $O(\dot J^2)$.}
The evolution equation for~$f$ acquires two contributions:
\begin{equation}\label{eq:app-fv}
  \partial_v f = -\varepsilon\,\ell^2\,\dot J\,f_{\rm s}
  \;-\; r\,\dot J^2 + O(\varepsilon^2\dot J)\,.
\end{equation}
The first term describes the adiabatic drift of the blackening function
(the AdS radius is changing); the second is the irreversible kinetic
energy deposited into the black hole, which integrates to
$\delta E_{\rm irr} \sim (\Delta J)^2/\tau \to 0$ as $\tau\to\infty$.

\paragraph{Order summary.}
\begin{center}
\begin{tabular}{@{}llll@{}}
\toprule
Order & Field & Formula & Origin \\
\midrule
$O(\varepsilon)$ & $\Phi_{\rm s}$, \eqref{eq:app-static-phi}
  & $\varepsilon\ell^2\ln(r/r_+)$ & radial running \\
$O(\varepsilon^2)$ & $\delta_{\rm s}$, \eqref{eq:app-static-delta}
  & $-\frac{1}{2}\varepsilon^2\ell^4\ln(r/R)$ & scalar backreaction \\
$O(\varepsilon^2)$ & $f_{\rm s}$, \eqref{eq:app-static-f}
  & running $\ell_{\rm eff}(r)$ & scalar backreaction \\
$O(\dot J)$ & $\psi$, \eqref{eq:app-psi}
  & horizon lag & finite response time \\
$O(\varepsilon\dot J)$ & $\Delta\delta$, \eqref{eq:app-delta-corr}
  & mixed lapse & slow-roll $\times$ adiabatic \\
$O(\dot J^2)$ & $\partial_v f$, \eqref{eq:app-fv}
  & irreversible heating & kinetic energy deposit \\
\bottomrule
\end{tabular}
\end{center}
Every correction vanishes at the endpoints ($\varepsilon = 0$,
$\dot J = 0$), so the geometry is exactly BTZ with $\ell_1$ at the
start and $\ell_2$ at the end.  During the interpolation, corrections
are controlled by two small parameters: $\varepsilon^2\ell^4\ln(R/r_+)$
(slow roll) and $1/(\omega_{\rm QNM}\tau)$ (adiabaticity), both of
which can be made arbitrarily small.  We have verified these scalings
numerically by solving the full nonlinear
system~\eqref{eq:app-C1}--\eqref{eq:app-W} with Chebyshev spectral
methods: the peak scalar deviation from the instantaneous static
solution scales as $\tau^{-1}$, consistent
with~\eqref{eq:app-psi}.

\subsection{Reference scheme for the AdS radius}
\label{app:scheme}

The static scalar EOM is solved by $\Phi_s'(r) = \varepsilon\ell^2/r$,
so the general profile is
\begin{equation}\label{eq:app-Phi-general}
  \Phi_s(r) = J_* + \varepsilon\ell^2\ln(r/r_*),
\end{equation}
where $r_*$ is an arbitrary reference radius and $J_* = \Phi_s(r_*)$
is the boundary value at that radius.  In an asymptotically AdS$_3$
background the leading boundary behaviour $\Phi_s \to \varepsilon\ell^2
\ln r + \mathrm{const}$ is the standard log running of a marginal
source: the coefficient $\varepsilon\ell^2$ is fixed by the potential
slope, while the additive constant is the renormalised source.  The
choice of $r_*$ is a renormalisation scheme.

\paragraph{Two natural schemes.}
\emph{Scheme A} (boundary BC): $r_* = R$ and $J_* = J \equiv \Phi_s(R)$.
The AdS radius is defined via the boundary potential value,
$V(J) = -1/\ell^2_A$.
\emph{Scheme B} (horizon BC): $r_* = r_+$ and $J_* = J_+ \equiv
\Phi_s(r_+)$.  The AdS radius is defined via the horizon potential
value, $V(J_+) = -1/\ell^2_B$.  The two scalar values are related by
$J_+ = J - \varepsilon\ell^2\ln(R/r_+)$, and the two definitions of
$\ell$ differ by
\begin{equation}\label{eq:app-ell-shift}
  \frac{1}{\ell^2_B} - \frac{1}{\ell^2_A}
  = V(J) - V(J_+) = \varepsilon^2\ell^2\,\ln(R/r_+).
\end{equation}

\paragraph{Origin of the shift: on-shell action.}
Using the trace of the Einstein equation,
$R - \tfrac{1}{2}(\partial\phi)^2 - 2V = 4V(\phi)$, so that the bulk
integrand on-shell is $4V(\Phi_s)$.  The static bulk action in
Scheme A evaluates to
\begin{equation}\label{eq:app-onshell-bulk}
  S_{\rm bulk}\big|_{r_+}^{R}
  = \frac{\beta}{2G}\!\left[
  \frac{V(J)}{2}(R^2-r_+^2)
  - \frac{\varepsilon^2\ell^2}{4}(R^2-r_+^2)
  + \frac{\varepsilon^2\ell^2\,r_+^2}{2}\ln(R/r_+)\right]
\end{equation}
(integrating $V(\Phi_s) = V(J) + \varepsilon^2\ell^2\ln(r/R)$ against
$\sqrt{-g}\approx r$).  The $R^2$ divergences are the usual gravity
divergences and are removed by Gibbons--Hawking and cosmological-
constant counterterms.  The new feature is the logarithmic divergence
$\propto r_+^2\ln(R/r_+)$ sourced by the scalar; this is the standard
log anomaly of a marginal source, removed by a counterterm of the form
\begin{equation}\label{eq:app-logct}
  S_{\rm ct,\,log}
  = -\frac{\beta\,r_+^2\,\varepsilon^2\ell^2}{8G}\,\ln(R/\mu),
\end{equation}
where $\mu$ is an arbitrary renormalisation scale.  Choosing $\mu = r_+$
absorbs the log directly, leaving a finite renormalised action.

\paragraph{Effect on $f_s$.}
The same scheme choice propagates to $f_s$.  Integrating
constraint~\eqref{eq:app-C2} in Scheme A with
$V(\Phi_s) = -1/\ell^2_A + \varepsilon^2\ell^2\ln(r/R)$ gives
\begin{equation}\label{eq:app-fs-schemeA}
  f_s^{(A)}(r)
  = \frac{r^2-r_+^2}{\ell^2_A}
  - \varepsilon^2\ell^2\bigl[(r^2 - r_+^2/2)\ln(r/r_+)
  - (r^2-r_+^2)/4 + (r^2-r_+^2)\ln(r_+/R)\bigr].
\end{equation}
Substituting~\eqref{eq:app-ell-shift} converts this to
\begin{equation}\label{eq:app-fs-schemeB}
  f_s^{(B)}(r)
  = \frac{r^2-r_+^2}{\ell^2_B}
  - \varepsilon^2\ell^2\bigl[(r^2 - r_+^2/2)\ln(r/r_+)
  - (r^2-r_+^2)/4\bigr],
\end{equation}
which is~\eqref{eq:app-static-f}.  The $\ln(R/r_+)$ piece has been
absorbed into the redefinition of $\ell$.  The bulk geometry is
identical; only the identification of ``the'' AdS radius differs
between schemes.

\paragraph{Why Scheme B is natural.}
With $\ell$ defined at the horizon, every horizon-thermodynamic
quantity ($T_H = r_+/(2\pi\ell_B^2)$, $S = 2\pi r_+/(4G)$, the
microcanonical energy, etc.) is cutoff-independent.  Scheme A would
make $\ell$ run with $R$, polluting every horizon observable with
log-cutoff dependence.  The boundary
condition~\eqref{eq:app-static-phi} should therefore be read as
\emph{$\Phi_s(r_+) = J$ at the horizon}, with $\ell$ defined by
$V(J) = -1/\ell^2$, and the boundary value $\Phi_s(R)$ is then a
derived quantity, $\Phi_s(R) = J + \varepsilon\ell^2\ln(R/r_+)$,
divergent as $R \to \infty$.  All the formulas in the remainder of
the appendix are consistent with this scheme.

In the time-dependent Lorentzian interpolation of
Section~\ref{sec:btz}, the natural choice is the instantaneous
Scheme B: $\ell(t)^2 = -1/V(\Phi(r_+,t))$, locking the AdS scale to
the slowly varying horizon value of the scalar.  It is this $\ell(t)$
that appears in the common-frame instantaneous energy
$E(t) = r_+^2/(8G\ell(t))$ and in the smoothness angles
$\Delta\theta_i = \beta r_+/\ell_i$ of~\eqref{eq:contour-angles}.

\section{Single-microstate insertion: FZZT brane in JT gravity}
\label{app:fzzt}

We illustrate the spectral formula~\eqref{eq:W2-leading} with a
worked example: a single-microstate perturbation
$\delta\rho(E) = \varepsilon\,\delta(E - E_*)$ realised gravitationally
through an FZZT brane in the JT gravity matrix
integral~\cite{Saad:2019}.

\paragraph{Caveat on order of operations.}
In the matrix-integral formulation $\rho(E)$ is itself a random
object, and the spectral formula
\begin{equation}\label{eq:app-fzzt-W2}
  \mathcal{W}^2 = \beta^2\!\int \frac{[\Delta n(E)]^2}{\rho_1(E)}\,
  \ee^{-2\beta E}\,\dd E
\end{equation}
contains $1/\rho_1$ in the integrand.  The fully correct ensemble
average is
\begin{equation}\label{eq:app-fzzt-correct}
  \big\langle \mathcal{W}^2\big\rangle_{\rm ens}
  = \beta^2 \int \Big\langle \frac{[\Delta n(E)]^2}{\rho_1(E)}
  \Big\rangle_{\rm ens}\,\ee^{-2\beta E}\,\dd E,
\end{equation}
which requires connected correlators of $\Delta n$ with $1/\rho$
and has no clean diagrammatic expansion in the matrix model (the
$1/\rho$ moments mix all orders in $\ee^{-S_0}$).
What we compute below instead is the simpler
\begin{equation}\label{eq:app-fzzt-simplified}
  \mathcal{W}^2_{\rm avg} \;\equiv\; \beta^2\!\int
  \frac{[\langle\Delta n\rangle_{\rm ens}]^2}{\langle\rho_1
  \rangle_{\rm ens}}\,\ee^{-2\beta E}\,\dd E,
\end{equation}
i.e.\ we use the ensemble-averaged spectral density and CDF
perturbation, then plug into~\eqref{eq:app-fzzt-W2}.  This is the
leading-order semiclassical answer; the discrepancy between
$\langle\mathcal{W}^2\rangle_{\rm ens}$
and~$\mathcal{W}^2_{\rm avg}$ lives at $O(\ee^{-S_0})$ and would
require the full connected moment calculation, which we do not
attempt here.

\subsection{FZZT brane as a microstate insertion}

A fractional eigenvalue insertion at $E_*$ is implemented by
\begin{equation}\label{eq:app-fzzt-det}
  \det(H - E_*)^{2\varepsilon} \;=\; \ee^{\varepsilon F},
  \qquad
  F \;=\; \sum_j \log(\lambda_j - E_*)^2.
\end{equation}
The modified saddle equation
$V'(\lambda_i) = 2\sum_{j\ne i}(\lambda_i - \lambda_j)^{-1} +
2\varepsilon/(\lambda_i - E_*)$ shows that the insertion adds
$\varepsilon$ eigenvalues at $E_*$ to the spectrum.

The ensemble-averaged density is
\begin{equation}
\rho_2 = \frac{\langle e^{\varepsilon F}\rho_1\rangle }{\langle e^{\varepsilon F}\rangle}
\end{equation}
where the division by $\langle e^{\varepsilon F}\rangle$ is performed to normalise the ensemble average.
Expanding in powers of $\varepsilon$ the
disconnected piece $\langle F\rangle\langle\rho\rangle$ cancels
exactly against the normalisation $\langle \ee^{\varepsilon F}\rangle$,
leaving only the connected cylinder contribution.  Using the
SSS identity $F = -2\int_0^\infty (\dd\beta'/\beta')\,
\ee^{-\beta'\kappa^2}\,Z(\beta')$ with $\kappa^2 = -E_*$, taking
$\kappa \to i\sqrt{E_*}$ to analytically continue inside the
spectrum, and extracting the real part via
Sokhotski--Plemelj~\cite{Saad:2019}, one finds
\begin{equation}\label{eq:app-fzzt-deltarho}
  \langle\delta\rho(E)\rangle_{\rm ens} \;=\; \varepsilon\,
  \delta(E - E_*) \;+\; O(\ee^{-S_0}),
\end{equation}
confirming~\eqref{eq:app-fzzt-det} at the level of the averaged
spectrum.

\subsection{The Boltzmann--Wasserstein distance}

The CDF perturbation is $\langle\Delta n(E)\rangle_{\rm ens} =
\varepsilon\,\theta(E - E_*)$, so~\eqref{eq:app-fzzt-simplified}
gives
\begin{equation}\label{eq:app-fzzt-W2-explicit}
  \mathcal{W}^2_{\rm avg} \;=\; \beta^2\,\varepsilon^2
  \int_{E_*}^{\infty} \frac{\ee^{-2\beta E}}{\rho_1(E)}\,\dd E.
\end{equation}
The integral runs from $E_*$ upward: the inserted state has
created a step in the CDF that contributes to the BW distance only
at energies above $E_*$.  The integrand combines two competing
effects --- thermal suppression $\ee^{-2\beta E}$ and the
transport cost $1/\rho_1(E)$ (sparse spectra cost more per
inserted state).

The matrix-integral steps above (Vandermonde insertion, cylinder
factorisation, Sokhotski--Plemelj continuation) use only the
universal double-scaled structure and apply to any matrix model
with a bulk dual; only the specific form of $\rho_0(E)$ in the
denominator of~\eqref{eq:app-fzzt-W2-explicit} is model-specific.
For concreteness we evaluate it for JT gravity, with
$\rho_0(E) = \ee^{S_0}\sinh(2\pi\sqrt{2E})/(4\pi^2)$.  The integral
is dominated by its lower endpoint for $E_* \gtrsim 1/(2\beta)$,
giving
\begin{equation}\label{eq:app-fzzt-large}
  \mathcal{W}^2_{\rm avg} \;\approx\; \frac{\beta\,\varepsilon^2\,
  \ee^{-2\beta E_*}}{\rho_0(E_*)}, \qquad E_* \gg 1/(2\beta).
\end{equation}
The distance is doubly suppressed: Boltzmann
suppression~$\ee^{-2\beta E_*}$ for a state thermally inaccessible
at temperature $1/(2\beta)$, and density suppression
$1/\rho_0(E_*)$ for a single insertion being a negligible fraction
of an exponentially large bulk spectrum.  At the spectral edge
$E_* \to 0$, the integral saturates at a finite
$O(\ee^{-S_0})$ value depending only on~$\beta$ --- the inserted
state is maximally distinctive against the sparse near-edge
spectrum but not Boltzmann-suppressed.

In this case the difference between the two CDF's
has a geometric interpretation. The difference in number of states
is computed by a geometry that corresponds to a disc
with a hole punched in the middle by an FZZT brane. 
The hole represents the fact that the difference $\Delta n$
counts no states below $E_*$. Cutting this punctured disc along $\tau = 0$
slice (the Hartle--Hawking preparation of the corresponding double sided state) 
yields a bulk Cauchy slice severed by the brane: the two asymptotic sides are no 
longer connected through the interior. The explicit microstate has cut the 
ensemble-averaged wormhole calculating the $\Delta n$.

\section{Derivation of the four-point representation}
\label{app:4pt}

This appendix records the perturbative analysis of the BW distance at
$O(\varepsilon^4)$ that underlies the four-point
representation~\eqref{eq:W2-otoc} stated in
Section~\ref{sec:pert-otoc}.  We first establish the all-orders
diagonal form of the matrix $X = \ee^{-\beta H_1} -
U_{\rm com}\ee^{-\beta H_2}U_{\rm com}^\dagger$ in the eigenbasis of
$H_1$ (\S\ref{app:4pt-diag}), expand it to obtain the
$O(\varepsilon^4)$ piece for a primary perturbation
(\S\ref{app:4pt-expand}), and convert the resulting
$(\delta_2 E_n)^2$ structure into the time-averaged thermal four-point
representation~\eqref{eq:W2-otoc} (\S\ref{app:4pt-derive}).

\subsection{Diagonality of $X$ at all orders}
\label{app:4pt-diag}

The starting point is a structural observation that, in hindsight,
follows directly from the BW formula~\eqref{eq:W-ell2}: the comonotone
unitary diagonalises $\ee^{-\beta H_2}$ in the eigenbasis of~$H_1$.
Indeed, $U_{\rm com}|n_\varepsilon\rangle = |n\rangle$ where
$|n_\varepsilon\rangle$ is the eigenstate of $H_2 = H_1 + \varepsilon V$
obtained from $|n\rangle$ by smooth continuation in~$\varepsilon$, so
\begin{equation}\label{eq:Ucom-diag}
  U_{\rm com}\,\ee^{-\beta H_2}\,U_{\rm com}^\dagger
  = \sum_n |n\rangle\,\ee^{-\beta E_n^{(\varepsilon)}}\,\langle n|,
\end{equation}
which is diagonal at \emph{all} orders in~$\varepsilon$ in the
$H_1$-eigenbasis.  The matrix $X$ is therefore diagonal too:
\begin{equation}\label{eq:X-all-orders}
  X_{nn} = \ee^{-\beta E_n} - \ee^{-\beta E_n^{(\varepsilon)}}\,,
  \qquad
  \mathcal{W}^2 = \Tr X^2
  = \sum_n\bigl(\ee^{-\beta E_n} - \ee^{-\beta E_n^{(\varepsilon)}}\bigr)^{\!2}.
\end{equation}
This is the discrete BW formula~\eqref{eq:W-ell2} for the sorted
Boltzmann weights: the off-diagonal $V_{mn}$ structure that appeared
in the derivation of $X$ at $O(\varepsilon)$ is fully absorbed into
the comonotone~$U_{\rm com}$.

The identification $U_{\rm com}|n_\varepsilon\rangle = |n\rangle$ uses
the smooth continuation of eigenstates in $\varepsilon$, while the
rigorous BW formula~\eqref{eq:W-ell2} pairs
eigenvalues by \emph{rank} --- the $k$-th smallest of $H_1$ with the
$k$-th smallest of $H_2$.  These two pairings agree if and only if
smooth eigenstate continuation preserves rank, i.e.\ no level
crossings occur along the family $H_1 + s V$ for $s \in [0,\varepsilon]$.
Three observations make this safe:

(i) For a non-degenerate spectrum and a generic perturbation $V$, level
crossings are codimension-two phenomena (von~Neumann--Wigner) and are
avoided by the one-parameter flow.

(ii) Second-order perturbation theory contributes a
nearest-neighbour level-repulsion piece,
$\bigl(\delta_2 E_{n+1} - \delta_2 E_n\bigr) \supset 2|V_{n,n+1}|^2/(E_{n+1}-E_n) > 0$,
which widens adjacent gaps under the second-order shift.

(iii) The regime where two levels could swap,
$\varepsilon |V_{n,n+1}| \gtrsim |E_n - E_{n+1}|$, coincides with the
regime where the perturbation series for $\delta E_n$ stops converging.
The formula is internally consistent: it is valid precisely where its
inputs are computable.

\subsection{Expansion of the energy shift}
\label{app:4pt-expand}

Writing $\delta E_n = E_n^{(\varepsilon)} - E_n = \varepsilon V_{nn} +
\varepsilon^2 \delta_2 E_n + O(\varepsilon^3)$ with
\begin{equation}\label{eq:delta2-En}
  \delta_2 E_n = \sum_{m\neq n}\frac{|V_{mn}|^2}{E_n - E_m}\,,
\end{equation}
and expanding~\eqref{eq:X-all-orders}, the BW distance reads
\begin{equation}\label{eq:W2-orders}
  \mathcal{W}^2 = \varepsilon^2\beta^2\!\sum_n V_{nn}^2\,\ee^{-2\beta E_n}
  + \varepsilon^4\beta^2\!\sum_n (\delta_2 E_n)^2\,\ee^{-2\beta E_n}
  + \bigl(\text{terms with at least one factor of }V_{nn}\bigr)
  + O(\varepsilon^5)\,.
\end{equation}
For a perturbation with vanishing thermal one-point function
(\S\ref{sec:pert-otoc}), the diagonal matrix elements $V_{nn}$ are
$O(\ee^{-S/2})$ ---
or vanish identically when a symmetry forbids them --- so the
$O(\varepsilon^2)$ term and the $V_{nn}$ cross terms are exponentially
suppressed in the entropy and the leading semiclassical contribution comes
from $O(\varepsilon^4)$:
\begin{equation}\label{eq:W2-eps4}
  \mathcal{W}^2 \approx \varepsilon^4\beta^2
  \sum_n (\delta_2 E_n)^2\,\ee^{-2\beta E_n}\,.
\end{equation}

\subsection{Four-point representation}
\label{app:4pt-derive}

Using the $i\epsilon$-prescription
\begin{equation}\label{eq:app-resolvent}
  \frac{1}{E_n - E_m} = -i\!\int_0^\infty\!\!\dd t\;
  \ee^{i(E_n - E_m)t - \eta t}\,,\qquad \eta\to 0^+,
\end{equation}
the second-order shift becomes
\begin{equation}\label{eq:app-delta2-time}
  \delta_2 E_n = -i\!\int_0^\infty\!\!\dd t\;\ee^{-\eta t}\,
  \langle n|V\,V(-t)|n\rangle,
\end{equation}
where $V(t) = \ee^{iHt}V\ee^{-iHt}$ and the $m=n$ piece is removed by
the $m\neq n$ restriction (automatic for primaries with $V_{nn} = 0$).

\paragraph{Diagonal projection identity.}
Squaring~\eqref{eq:app-delta2-time} requires evaluating
$\langle n|A|n\rangle\langle n|B|n\rangle$ at each eigenstate, which is
implemented via the identity
\begin{equation}\label{eq:app-diag-proj}
  \langle n|A|n\rangle\,\langle n|B|n\rangle
  = \lim_{T\to\infty}\frac{1}{T}\!\int_0^T\!\!\dd s\;
  \langle n|A\,B(s)|n\rangle\,,
  \qquad B(s) = \ee^{iHs}B\ee^{-iHs}\,,
\end{equation}
which holds in a non-degenerate spectrum: the time average over $s$
kills off-diagonal phases $\ee^{i(E_n - E_k)s}$ for $k\neq n$ and
leaves only the diagonal $k = n$ term. With degeneracies the same average instead retains all pairs with
$E_k = E_n$, projecting onto energy blocks rather than individual states;
read in the adapted basis of Section~\ref{sec:pert-ucom}, where $V$ is
block-diagonal at each energy, the formulae below hold unchanged.

\paragraph{Assembling.}
Squaring~\eqref{eq:app-delta2-time}, applying~\eqref{eq:app-diag-proj}
to project onto the diagonal in $|n\rangle$, and summing against
$\ee^{-2\beta E_n}/Z$, one obtains
\begin{equation}\label{eq:app-W2-4pt}
  \mathcal{W}^2 = -\varepsilon^4\beta^2
  \lim_{T\to\infty}\frac{1}{T}\!\int_0^T\!\!\dd s\!\int_0^\infty\!\!\dd t_1\!\int_0^\infty\!\!\dd t_2\;
  \ee^{-\eta(t_1+t_2)}\,
  \Tr\!\,\bigl(V(0)\,V(-t_1)\,V(s)\,V(s-t_2)\,\ee^{-2\beta H}\bigr) + O(\varepsilon^5)\,,
\end{equation}
which is~\eqref{eq:W2-otoc}.

\section*{Acknowledgements}

It is a pleasure to thank Claude (Anthropic) for extensive collaboration
throughout this project --- for working through calculations, catching
errors, helping with the exposition, writing and debugging the numerical
code, producing the figures, and being a patient and engaging discussion
partner. The project was a lot more fun than it would have been alone.

\end{document}